# The Solar and Geomagnetic Storms in May 2024: A Flash Data Report

Hisashi Hayakawa (1 – 4)*, Yusuke Ebihara (5 – 6), Alexander Mishev (7 – 8), Sergey Koldobskiy (7 – 8), Kanya Kusano (1), Sabrina Bechet (9), Seiji Yashiro (10 – 11), Kazumasa Iwai (1), Atsuki Shinbori (1), Kalevi Mursula (8), Fusa Miyake (1), Daikou Shiota (1, 13), Marcos V. D. Silveira (14), Robert Stuart (15), Denny M. Oliveira (11 – 12), Sachiko Akiyama (10 – 11), Kouji Ohnishi (16), Vincent Ledvina (17), Yoshizumi Miyoshi (1)

(1) Institute for Space-Earth Environmental Research, Nagoya University, Nagoya, Japan

(2) Institute for Advanced Research, Nagoya University, Nagoya, Japan

(3) Space Physics and Operations Division, RAL Space, Science and Technology Facilities Council, Rutherford Appleton Laboratory, Harwell Oxford, Didcot, UK

(4) Astro-Glaciology Laboratory, Riken Nishina Centre, Wako, Japan

(5) Research Institute for Sustainable Humanosphere, Kyoto University, Uji, Japan

(6) Unit of Synergetic Studies for Space, Kyoto University, Kyoto, Japan

(7) Sodankylä Geophysical Observatory, University of Oulu, Oulu, Finland

(8) Space Physics and Astronomy Research Unit, University of Oulu, Oulu, Finland

(9) Royal Observatory of Belgium, Brussels, Belgium

(10) The Catholic University of America, Washington, DC, USA

(11) NASA Goddard Space Flight Center, Greenbelt, MD, USA

(12) Goddard Planetary Heliophysics Institute, University of Maryland, Baltimore County, Baltimore, MD, USA

(13) National Institute of Information and Communications Technology, Japan

(14) University of São Paulo, Lorena School of Engineering, Lorena, Brazil

(15) Newtown Astronomical Society, Rhayader, UK

(16) Center for Liberal Art, Department of Engineering, National Institute of Technology, 716 Tokuma, Nagano, 3818550, Japan

(17) Geophysical Institute, University of Alaska Fairbanks, Fairbanks, AK, USA

* hisashi@nagoya-u.jp

**Abstract**

In May 2024, the scientific community observed intense solar eruptions that resulted in a great geomagnetic storm and auroral extension, highlighting the need to document and quantify these events. This study mainly focuses on their quantification. The source active region (AR 13664) evolved from 113 to 2761 millionths of the solar hemisphere between 4 May and 14 May. AR 13664's magnetic free energy surpassed $10^{33}$ erg on 7 May, triggering 12 X-class flares on $8 - 15$ May. Multiple interplanetary coronal mass ejections (ICMEs) were produced from this AR, accelerating solar energetic particles toward Earth. According to satellite and interplanetary scintillation data, at least 4 ICMEs erupted from 13664 eventually overcoming each other and combining. The shock arrival at 17:05 UT on 10 May significantly compressed the magnetosphere down to $\approx 5.04\ R_{\oplus}$, and triggered a deep Forbush Decrease. GOES satellite data and ground-based neutron monitors confirmed a ground-level enhancement from 2 UT to 10 UT on 11 May 2024. The ICMEs induced exceptional geomagnetic storms, peaking at a Dst index of $-412$ nT at 2 UT on 11 May, marking the sixth-largest storm since 1957. The AE and AL indices showed great auroral extensions that located the AE/AL stations into the polar cap. We gathered auroral records at that time and reconstructed the equatorward boundary of the visual auroral oval to 29.8° invariant latitude. We compared naked-eye and camera auroral visibility, providing critical caveats on their difference. We also confirmed global enhancements of storm-enhanced density of the ionosphere.

## 1. Introduction

Sunspot active regions occasionally direct interplanetary coronal mass ejections (ICMEs) to the Earth, disrupting near-Earth space environments, altering cosmic-ray radiation levels, perturbing terrestrial magnetic fields, and extending the auroral oval equatorward (Pulkkinen, 2007; Temmer, 2021; Cliver *et al.*, 2022; Kusano, 2023; Usoskin, 2023). Empirical evidence suggests that such solar eruptions occur more often in larger solar cycles, especially between their maximum and declining phase, although the maximum magnitude of the solar eruption is not necessarily dependent on the magnitude of the host solar cycle (Kilpua *et al.*, 2015; Lefèvre *et al.*, 2016; Chapman *et al.*, 2020; Owens *et al.*, 2022; Chapman and Dudok de Wit, 2024). Such solar eruptions occasionally cause significant terrestrial impacts such as extreme geomagnetic storms and the increase of charged particle flux throughout the atmospheric profile, leading to increased countrate of ground-based particle radiation detectors. These increases are known as ground level enhancements (GLEs; Cliver *et al.*, 2022; Usoskin *et al.*, 2023). Extending analyses to the extremity of solar eruptions and their terrestrial consequences is essential to bridge modern scientific data with past supersized events, develop statistical analyses, and understand impacts on terrestrial infrastructure (Daglis, 2001; Baker *et al.*, 2008; Beggan *et al.*, 2013; Riley *et al.*, 2018; Miyake *et al.*, 2019; Desai *et al.*, 2020; Hapgood *et al.*, 2021; Kusano, 2023; Usoskin *et al.*, 2023).

In this context, the last solar cycle, Solar Cycle 24 (from December 2008 to December 2019), was notably quiet in terms of disturbances in the solar-terrestrial environments (Watari, 2017; Manu *et al.*, 2022). This solar cycle marked the quietest sunspot number in the last century since Solar Cycle 14 in January 1902 – July 1913 (Clette *et al.*, 2023). Solar Cycle 24 witnessed up to X14.8 in September 2017 in the revised scale (Hudson *et al.*, 2024), geomagnetic storms reached a minimum Dst = −234 nT in March 2015 (Gopalswamy *et al.*, 2015, 2024; Liu *et al.*, 2015; Piersanti *et al.*, 2017; Webb and Nitta, 2017; WDC for Geomagnetism at Kyoto *et al.*, 2015), and hosted only two GLEs in May 2012 and September 2017 (Mishev *et al.*, 2014, 2018; Usoskin *et al.*, 2020).

The current solar cycle, Solar Cycle 25, commenced in December 2019. Predictions suggest that this cycle may be either as quiet as or slightly more active than Solar Cycle 24 (Pesnell and Schatten, 2018; Bhowmik and Nandi, 2018; McIntosh *et al.*, 2020; Mursula, 2023; Upton and Hathaway, 2023; Clette *et al.*, 2024), whereas lessons from the previous cycle caution us to closely monitor how these forecasts align with the actual temporal evolution of the sunspot number (Pesnell, 2016, 2020). Despite these predictions, Solar Cycle 25 has shown increased activity compared to Solar Cycle 24 in terms of disturbances in terrestrial environments (Oliveira *et al.*, 2024). Even prior to reaching its maximum, Solar Cycle 25 has already witnessed one severe geomagnetic storm (with a minimum Dst below −200 nT according to the classic classification of Loewe and Prölss (1997)) in April 2023, reaching a minimum Dst of −213 nT (Li and Jin, 2024), and one GLE in October 2021 (Mishev *et al.*, 2022; Papaioannou *et al.*, 2022; Klein *et al.*, 2022).

In May 2024, Solar Cycle 25 witnessed the emergence of a significant sunspot active region, triggering multiple solar eruptions and leading to a GLE and a substantial geomagnetic storm. Notably, this geomagnetic disturbance reached −412 nT in the real-time Dst index, which is classified as a "great" storm (min Dst < −350 nT) according to Loewe and Prölss (1997) and ranked as the sixth greatest in the Dst index starting from 1957 and the ninth greatest in the Dxt index starting from 1932 (WDC for Geomagnetism at Kyoto *et al.*, 2015; Karinen and Mursula, 2005; Mursula *et al.*, 2023). This geomagnetic storm extended auroral displays to lower magnetic latitudes (*e.g.*, Ghosh, 2024; Greshko, 2024; Karan *et al.*, 2024). This solar and geomagnetic storm has attracted the scientific community to start analyses on several aspects such as their solar radio bursts (Kruparova *et al.*, 2024), geomagnetic impacts in the Kp index and HPO index (Yamazaki *et al.*, 2024), local geomagnetic and ionospheric impacts around Italy (Spogli *et al.*, 2024), auroral imageries, and resultant satellite drags (Parker and Linares, 2024). Obviously more to come.

This solar-terrestrial storm serves as a valuable reference point for connecting modern scientific data with past great storms, particularly regarding geomagnetic impacts. Therefore, it is valuable to have a flash global overview of this solar and geomagnetic storm on the basis of what is available in the immediate aftermath. Here, we quantify observational datasets from this solar-terrestrial storm to facilitate further analysis and discussions regarding this exceptional event and to contextualise it with historical great storms.

## 2. The Source Solar Eruptions

This event occurred in an ascending phase – or near the maximum – of SC25. The daily sunspot number (Clette *et al.*, 2023) significantly increased on 2 May, reaching 187 on 6 May and 186 on 12 May. At the time of writing, some predictions place the expected maximum from late 2024 to early 2025 (McIntosh *et al.*, 2021; Upton *et al.*, 2023; Clette *et al.*, 2024).

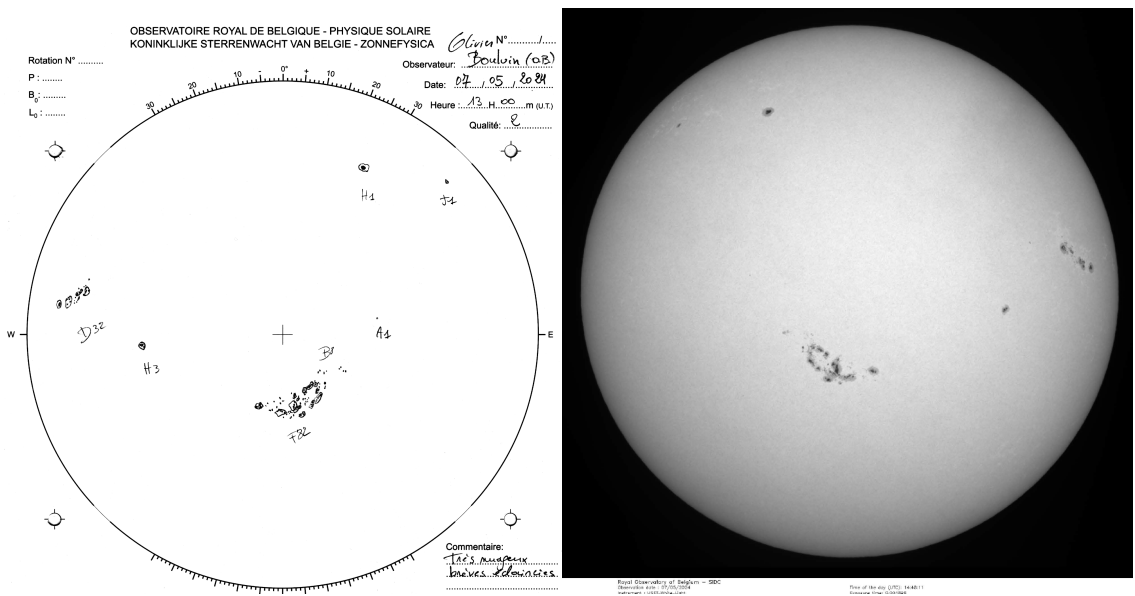

Figure 1: Sunspot drawing and white-light image on 07 May 2024, as recorded with Uccle Solar Equatorial Table telescopes in the Royal Observatory of Belgium (in a projected image). NOAA 13664 is located close to the central meridian.

A new active region (AR), known as AR 13664 according to the NOAA/SWPC region number, came into the south-eastern limb of the visible side of the solar disk on 1 May. Its temporal and spatial evolutions were documented in sunspot drawings and images in several observatories such as the Royal Observatory of Belgium using the Uccle Solar Equatorial Table (USET) telescopes (Bechet and Clette, 2022). This sunspot group started as a bipolar group, with two penumbrae and some intermediate sunspots in between. This active region developed rapidly from 3–7 May, and crossed the central meridian on 7 May. The area measured from the USET sunspot drawing was 1312 millionths of the visible solar hemisphere (hereafter MSH), based on the measurements with the Digi Sun software (*e.g.*, Clette, 2011; Hayakawa *et al*., 2023c). This group was still growing and developed into a Fkc McIntosh type, consisting of several moderately sized penumbral sunspots. This group continued to grow until 9 May up to an area of 2761 MSH, consisting of a compact penumbra containing most of the spots, as shown in the sunspot drawing and white-light image (Figure 1). The area remained relatively stable for a period before rotating over the limb on 14 May. The evolution of the group area, as measured from sunspot drawings, is illustrated in Figure 2, showing a rapid increase until 10 May.

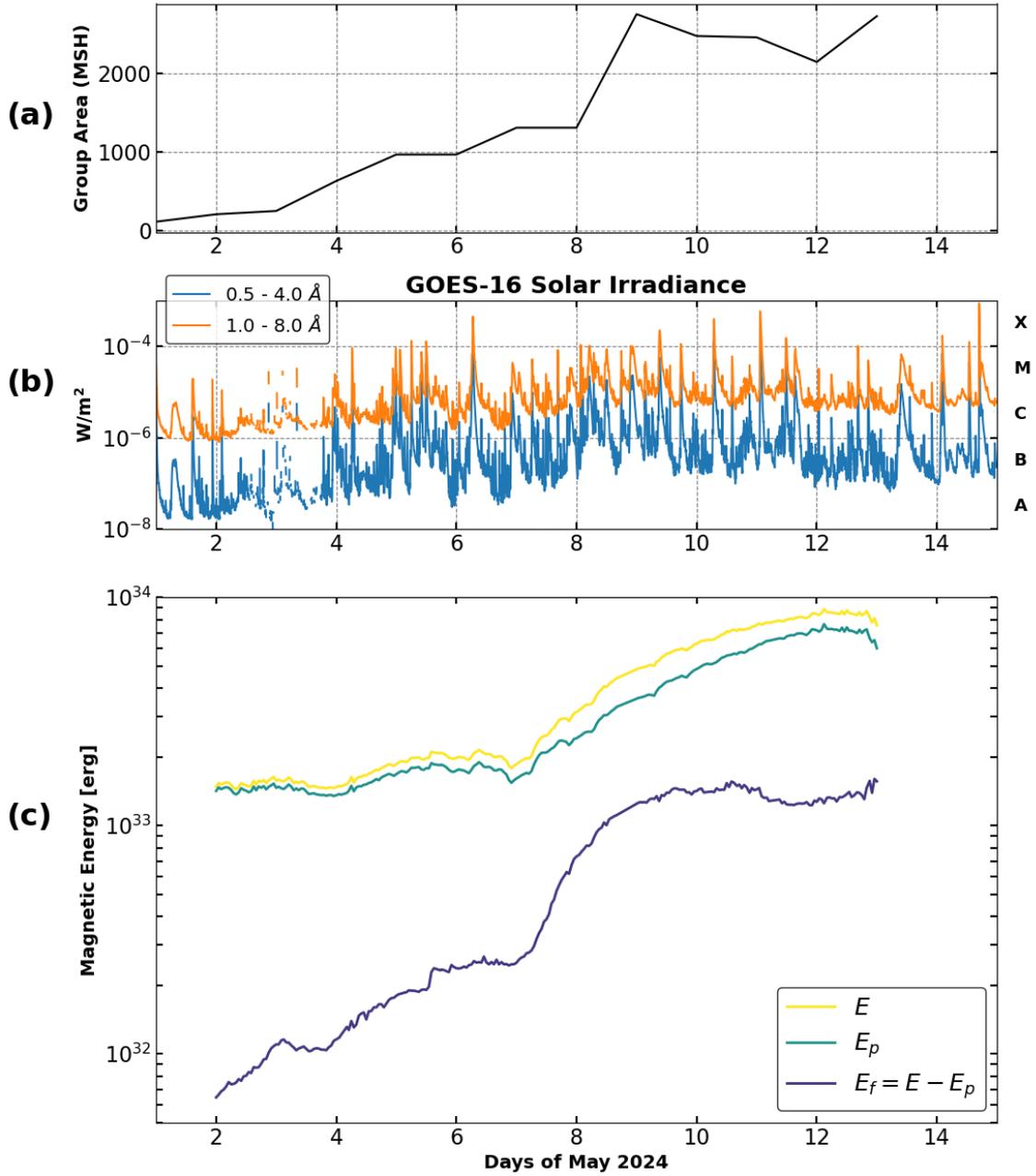

Figure 2: Temporal evolutions of the AR 13664 in millionths of the visible solar hemisphere (MSH) based on the USET measurements, GOES-16 X-ray fluence in 0.5–4.0 Å and 1.0–8.0 Å, as derived from the NOAA web portal, and the magnetic energy of the active region NOAA 13664 as a function of the day of May 2024. $E$, $E_p$, and $E_{free}$ are the total energy of the nonlinear force-free field, the potential magnetic field, and the free magnetic energy ($E - E_p$).

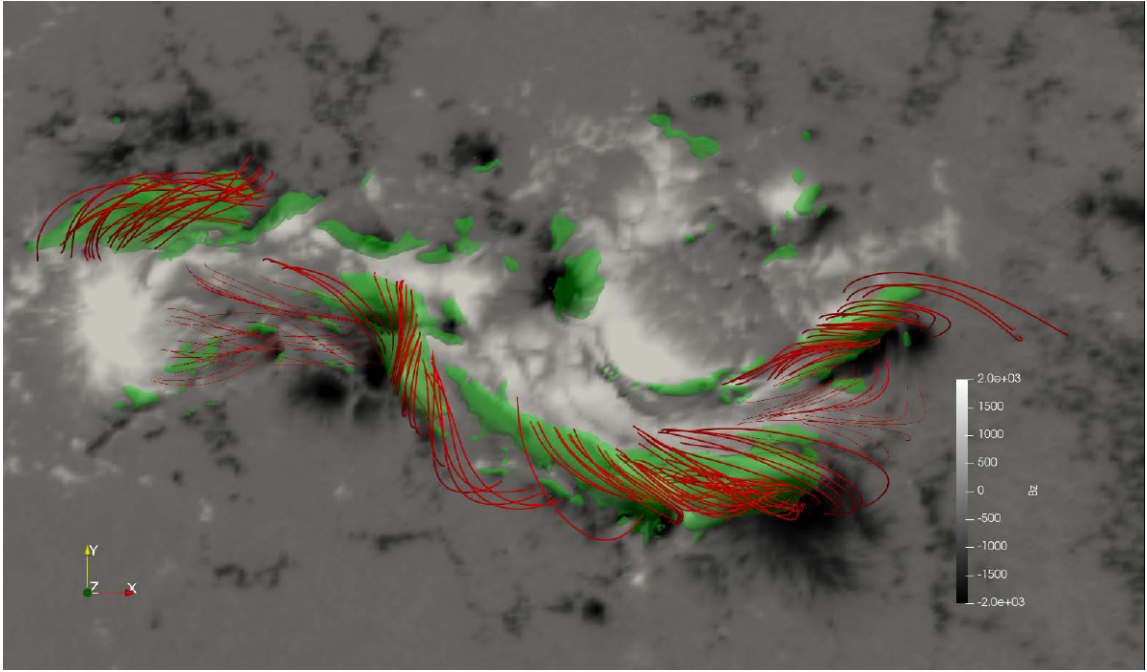

Figure 3: The distribution of the magnetic field of AR 13664 at 06:00 UT on 10 May 2024, just before the X3.9 flare. The grayscale background indicates the radial magnetic field (B$_r$ [G]), and the green surface is the contour of 10$^3$ G for the nonpotential field intensity, corresponding to HiFERs. Red lines show selected magnetic field lines of the nonlinear force-free field.

The unsigned magnetic flux constantly increased from 7 May to 1.9×10$^{23}$ Mx before it reached the west limb. We extrapolated the nonlinear force-free field (NLFFF) (Inoue *et al.*, 2014) of this AR using near-real-time data of the photospheric vector magnetic field from the Spaceweather HMI Active Region Patch (SHARP) near-real-time (NRT) dataset (Bobra *et al.*, 2014). Figure 2 shows the magnetic free energy, which is defined as the difference between the volume-integrated magnetic energy of the NLFFF and the potential magnetic field. We can see that the magnetic free energy increased before the M-class flares on 2 May and quickly grew on 7 May to ≈ 10$^{33}$ erg before the consecutive X-class flares (Figure 2 and Table 1). The maximum values of the unsigned magnetic flux and magnetic free energy are as large as those of AR NOAA 12192, which provides the largest sunspot area, unsigned flux, and magnetic free energy in SC 24 (Karimov *et al.*, 2024).

Figures 3 shows the distribution of the radial magnetic field and the contours of nonpotential magnetic field intensity |**B**-**B**$_p$|=1000 G (green surfaces), where **B** and **B**$_p$ are the magnetic fields from the SHARP-NRT data and the potential magnetic field. Red lines are the magnetic field lines of force for the nonlinear force-free field extrapolated by the magnetohydrodynamic (MHD) relaxation method (Inoue *et al.*, 2014). The notable feature of this region is the presence of several high free energy regions (HiFERs) where the nonpotential field intensity exceeds 1000 G. Likely, the large free energy and the complex structure of the HiFERs distribution may cause consecutive large flares (Kusano *et al*., 2020).

Eventually, this active region hosted 12 X-class flares from 1 May to 15 May (Table 1 and Figure 2). Most of these were produced during transit over the visible solar hemisphere, while four were produced after the AR went over the western limb onto the far side. This AR produced a flare as great as X8.7. Even though the flare partially occurred on the western limb, this was the largest solar flare in the current solar cycle (SC25) at the time of writing at that time. Obviously more to come in this solar cycle, as exemplified with the X9.0 flare from AR 13842 on 3 October. Figure 2 shows the soft X-ray flux continuously above the M1 level from 8 to 12 May.

Table 1: Summary of the X-class flares produced by AR 13664, with their peak time, GOES class, source location and associated CME (if any), on the basis of the SOHO/LASCO data calibration in August 2024. The flare magnitudes and source locations are derived from the Event Archives of Solar Software of the Lockheed Martin Solar & Astrophysics Laboratory with a minor modification for location of the limb eruptions. We have annotated the longitude of the limb flares on the western limb as WL, in contrast with the Event Archives of Solar Software. The CME (coronal mass ejection) speeds in km/s are derived from the SOHO/LASCO dataset that is in preparation under Nat Gopalswamy's team. Their travel time (in hh:mm) and arrival time are estimated under assumption of no deceleration of the CME speed in the fifth column as measured in the SOHO/LASCO data. This is an unrealistic assumption with a large error margin, as CMEs do not slow down much only in extreme preconditioning cases. We estimated travel time and arrival time for CMEs originated from the central meridian (< |45|°from the central meridian).

| ID | GOES peak time | GOES class | Location | CME speed | Travel time* | Arrival time* |
|---|---|---|---|---|---|---|
| 1 | 2024-05-08 05:09 | X1.0 | S22 W11 | 530 | 78:49 | 05-11 11:58 |
| 2 | 2024-05-08 21:40 | X1.0 | S20W17 | 952 | 43:53 | 05-10 17:33 |
| 3 | 2024-05-09 09:13 | X2.2 | S20 W24 | 1280 | 32:38 | 05-10 17:51 |
| 4 | 2024-05-09 17:44 | X1.1 | S17 W28 | 1024 | 40:48 | 05-11 10:32 |

| ID | GOES peak time | GOES class | Location | CME speed | Travel time* | Arrival time* |
|----|----------------|------------|----------|-----------|--------------|---------------|
| 1  | 2024-05-08 05:09 | X1.0 | S22 W11 | 530 | 78:49 | 05-11 11:58 |
| 2  | 2024-05-08 21:40 | X1.0 | S20W17 | 952 | 43:53 | 05-10 17:33 |
| 5  | 2024-05-10 06:54 | X3.9 | S17 W34 | 1006 | 41:31 | 05-12 00:25 |
| 6  | 2024-05-11 01:23 | X5.8 | S17 W44 | 1614 | 25:53 | 05-12 03:16 |
| 7  | 2024-05-11 11:44 | X1.5 | S19 W60 | No CME | -- | -- |
| 8  | 2024-05-12 16:26 | X1.0 | S20 W75 | No CME | -- | -- |
| 9  | 2024-05-14 02:09 | X1.7 | S17 WL | 805 | -- | -- |
| 10 | 2024-05-14 12:55 | X1.2 | S17 WL | 792 | -- | -- |
| 11 | 2024-05-14 16:51 | X8.7 | S18 WL | 2010 | -- | -- |
| 12 | 2024-05-15 08:37 | X3.4 | S18 WL | 1648 | -- | -- |

## 3. Coronal Mass Ejections

These solar eruptions of May 2024 released a considerable number of CMEs. From 5–15 May 2024, the Large Angle Spectroscopic Coronagraph (LASCO) observed at least 19 major CMEs (width ≥ 60°), including 10 halo CMEs, which were produced by AR 13664 according to the dataset being prepared by Gopalswamy's team. The calibrated LASCO data became available in August 2024 in the Coordinated Data Analysis Workshop (CDAW) for creating the CME Catalog (Yashiro *et al*., 2004; Gopalswamy, 2009).

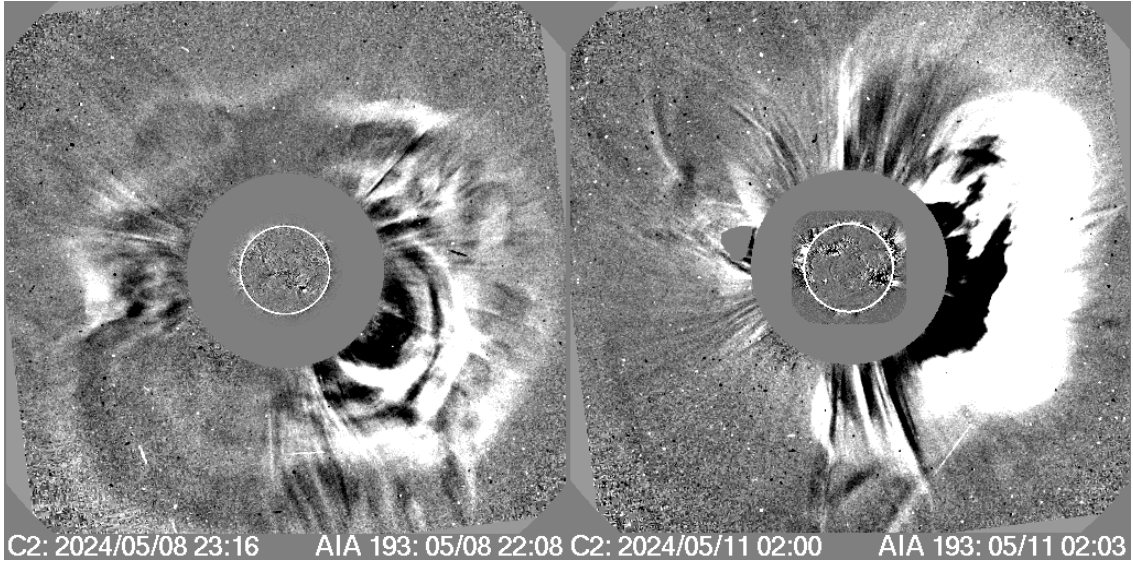

Figure 4: Examples of halo CMEs produced by AR 13664, as observed by LASCO C2 (Brueckner *et al.*, 1995) from 22:08–23:16 UT on 8 May 2024 (left) and 02:00–02:03 UT on 11 May 2024 (right). Image courtesy of the CDAW of NASA.

We have identified the CME that was most likely responsible for inducing the space environment disturbances observed near the Earth on 11 May with those comprising two flares (X1.0 at 21:40 UT and M9.8 at 22:27 UT) on 8 May that occurred in quick succession. The Solar Dynamics Observatory/Atmospheric Imaging Assembly (SDO/AIA) (Lemen *et al.*, 2012) observed the eruptive features in both flares. However, only a single halo CME was observed at 22:24 UT (#2 in Table 1; hereafter referred to as CME2), as shown in Figure 4. The two eruptions observed in the AIA likely merged before reaching the field of view of LASCO C2. The projected speed of CME2 was 952 km/s and if it travelled at this speed without any deceleration/acceleration, it would have reached Earth at ≈ 17:33 UT on May 10 (Table 1). While this is slightly later than the actual CME arrival of 17:05 UT on 10 May, this offset shows a considerably good agreement in contrast with the usual error margins (up to ±8 to ±12 hours) of the CME arrival time predictions (Gopalswamy *et al.*, 2001, 2005; Zhong *et al.*, 2024). In several cases, CMEs propagate earthward faster than expected from their apparent propagation speed measured in coronagraphs. In such cases, the actual arrival time comes earlier than the estimated arrival time. Based on these findings, we believe that CME2 was associated with the storm sudden commencement (SSC) at 17:05 UT on 10 May. This estimated arrival time indicated that CME2 likely stacked on CME1 (#1 in Table 1). Additionally, two more halo CMEs were observed, which were associated with M-class flares (Table 2) produced by the AR 13664. CME2 likely collided and stacked upon the three preceding CMEs originating from the same

AR, resulting in a complex ICME structure, as discussed later. Therefore, the great geomagnetic storm of 10/11 May 2024 is not associated with a single fast ICME but with a complex accumulation of multiple ICMEs and southward interplanetary magnetic fields (IMFs).

Table 2: Major CMEs produced by AR 13664, as observed by the Large Angle Spectroscopic Coronagraph (LASCO). These data were obtained from the dataset being prepared by Nat Gopalswamy's team. The table presents the dates, start times in Universal Time (UT), speed, and angular width (WD) of the CMEs, as well as the flare start times, GOES X-ray classes, location, and NOAA AR numbers for the solar sources. The values were updated after a calibration conducted in August 2024. We have annotated the limb eruption in the western limb as WL (western limb) to annotate their source location.

| Date | CME time (UT) | CME speed (km/s) | CME WD (°) | Flare time (UT) | Flare class (GOES class) | Location | NOAA |
|---|---|---|---|---|---|---|---|
| 2024-05-05 | 15:36 | 353 | 157 | 14:12 | C7.5 | S20E19 | 13664 |
| 2024-05-07 | 04:48 | 311 | 56 | 02:59 | C9.5 | S21E05 | 13664 |
| 2024-05-08 | 02:24 | 358 | 169 | 02:16 | M3.4 | S17W04 | 13664 |
| 2024-05-08 | 05:36 | 530 | 360 | 04:37 | X1.0 | S22W11 | 13664 |
| 2024-05-08 | 12:24 | 677 | 360 | 11:26 | M8.7 | S20W17 | 13664 |
| 2024-05-08 | 22:36 | 952 | 360 | 21:08 | X1.0 | S20W17 | 13664 |
| 2024-05-09 | 09:24 | 1280 | 360 | 08:45 | X2.2 | S20W26 | 13664 |
| 2024-05-09 | 12:24 | 563 | 53 | 11:52 | M3.1 | S16W40 | 13664 |
| 2024-05-09 | 19:18 | 1024 | 360 | 17:23 | X1.1 | S14W28 | 13664 |
| 2024-05-10 | 07:12 | 953 | 360 | 06:27 | X3.9 | S17W34 | 13664 |
| 2024-05-11 | 01:36 | 1614 | 360 | 01:10 | X5.8 | S15W45 | 13664 |
| 2024-05-11 | 16:12 | 846 | 93 | 14:46 | M8.8 | S15W49 | 13664 |

| 2024-05-13 | 09:12 | 1690 | 360 | 08:48 | M6.6 | S20W81 | 13664 |
| 2024-05-14 | 01:48 | 331  | 63  | 01:23 | M2.6 | S17W88 | 13664 |
| 2024-05-14 | 02:24 | 805  | 172 | 02:03 | X1.7 | S19W88 | 13664 |
| 2024-05-14 | 13:00 | 749  | 78  | 12:40 | X1.2 | S17W92 | 13664 |
| 2024-05-14 | 17:00 | 2010 | 360 | 16:46 | X8.7 | S18W96 | 13664 |
| 2024-05-15 | 08:36 | 1648 | 360 | 08:13 | X3.5 | S18W98 | 13664 |
| 2024-05-15 | 10:48 | 1024 | 194 | 09:47 | M3.6 | WL     | 13664 |
| 2024-05-15 | 21:18 | 1355 | 360 | 20:30 | C5.2 | WL     | 13664 |

Subsequently, the AR produced more halo CMEs, including two major flares (X1.1 at 09:24 on 9 May and X3.8 at 07:12 on 10 May) launched ICMEs at similar speeds (1226 and 1006 km/s) and another (X5.8 at 01:36 on 11 May) launched a faster ICME (1512 km/s), as shown in Figure 4. These ICMEs probably arrived on Earth at ≈ 9:00 UT on 12 May. This small difference between the observed and estimated arrival times, as indicated in Table 1, may imply that the preceding ICME sequence cleared the interplanetary space, thereby allowing subsequent ICMEs to propagate without significant deceleration (Liu *et al*., 2014, 2019; Shiota and Kataoka, 2016; Temmer *et al*., 2017; Desai *et al*., 2020). Additionally, the solar wind remained compressed owing to these ICMES, resulting in a prolonged geomagnetic storm.

CMEs are typically faster or slower than their measured projected velocity, owing to the propagation process in the interplanetary space and their actual 3D structure. This is because the measured projected velocity, which is based on transverse expansion speed, is not necessarily equal to the propagation speed in the Sun-Earth direction. The difference in the speeds depends on the shape of the CME. Therefore, the actual arrival time mostly varies from the estimated arrival time, as indicated in Table 1. Furthermore, AR 13664 produced multiple CMEs at significantly different speeds over several days. Therefore, the relationship between the CMEs observed through coronagraphs and the ICME structures observed in situ can be considerably more complex. To interpret this relationship, MHD (magnetohydrodynamic) simulations of multiple CME propagations should be further analysed. Our results are considerably different from Spogil *et al*. (2024), who used the CME speeds automatically obtained from CACTus (Robbrecht and Berghmans, 2004; Robbrecht *et al*., 2009) and associated the solar eruptions on 9–11 May with this great geomagnetic storm. The CACTus median speeds are generally lower than manual measurements and their maximum speeds are occasionally extremely high (*e.g.*, figure 14 of Robbrecht and Berghmans (2004)). Owing to their complexity, in-depth analyses of the interplay between ICMEs are currently underway by Daikou Shiota's team.

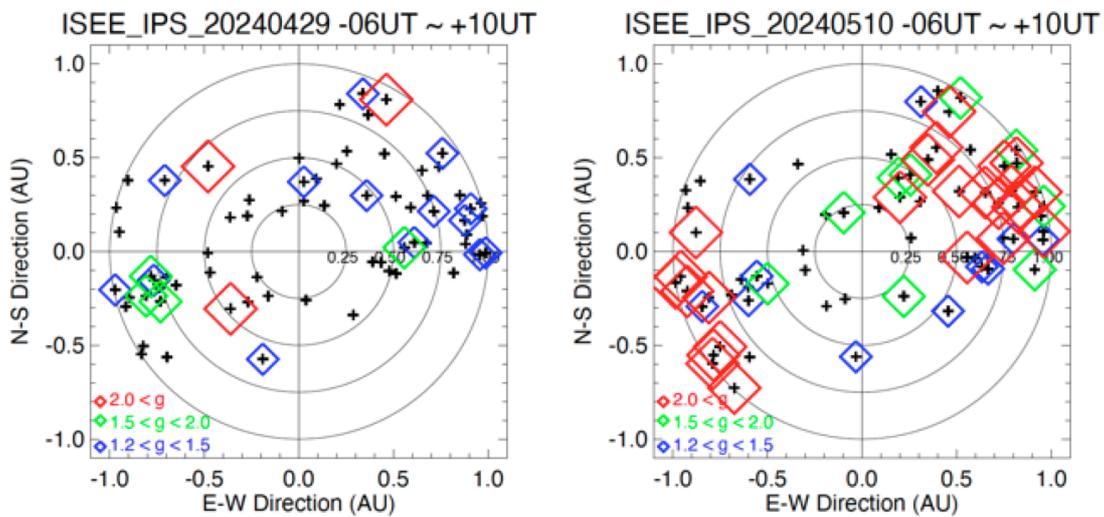

Figure 5: All-sky map of the IPS amplitude (g-value) observed by ISEE, Nagoya University on (a) 29 April 2024 when the AR 13663 was not on the disk and (b) 10 May 2024. +: all observed radio sources; diamonds: radio sources with g-values as follows: 2.0 < g (red), 1.5 < g < 2.0 (green) and 1.2 < g < 1.5 (blue).

Radio emissions from sources from the outside of the solar system can be scattered when ICMEs cross the line-of-sight during observation and are detected as interplanetary scintillations (IPS). IPS observations have been used to study ICMEs in the interplanetary space between the SOHO coverage and the geospace space satellites located around L1. In this study, we analysed the IPS observational data from early May measured by the ISEE, Nagoya University, at 327 MHz (Tokumaru *et al*., 2011). Immediately before the shock reached Earth, an increasing number of large-amplitude IPS responses were observed from 8–10 May, as shown in Figure 5. It is evident that large-amplitude IPS responses were detected across numerous radio sources, particularly in the line-of-sight in the direction of large solar separation angles, where these ICMEs were expected to be situated. IPS response enhancements can be associated with high-density regions where fast-propagating ICMEs accumulate background solar winds (Iwai *et al.*, 2019, 2021). Multiple ICME mergers can enhance the compression of the solar wind plasma (Scolini *et al.*, 2020). Notably, Koehn *et al.* (2022) tried a parametric study on this phenomenon and showed how two CMEs with velocity of 500 km/s and 1500 km/s could combine to produce a great geomagnetic storm of minimum Dst of ≈ −400 nT (figure 10 of Koehn *et al.* (2022)). This is similar to what happened in May 2024. Furthermore, ICME compression can generate intense radio scattering if a region of concentrated density disturbance is created (Equation 1 of Iwai *et al.* (2019); Young (1971)). This may support our aforementioned discussion of the ICME–ICME interactions. Further in-depth analyses are in preparation by Kazumasa Iwai's team.

## 4. Cosmic Ray Variations in Near-Earth Space and on the Ground

To monitor the cosmic ray intensity around this solar-terrestrial storm, we gathered proton flux data from the eastern and western monitors of the Solar and Galactic Proton Sensor (SGPS) onboard the GOES-16 satellite (Kress *et al.*, 2021) downloaded from the NOAA data portal. We also acquired ground-based neutron monitor records from high-latitude regions, namely those of Oulu (Oulu in Finland), THUL (Thule in Greenland), SOPO (South Pole in Antarctica), and MWSN (Mawson in Antarctica), from the Oulu neutron monitor website and the Neutron Monitor Database, as summarised in the International GLE Database (Usoskin, 2020). These data are shown in Figure 6 with their source URLs.

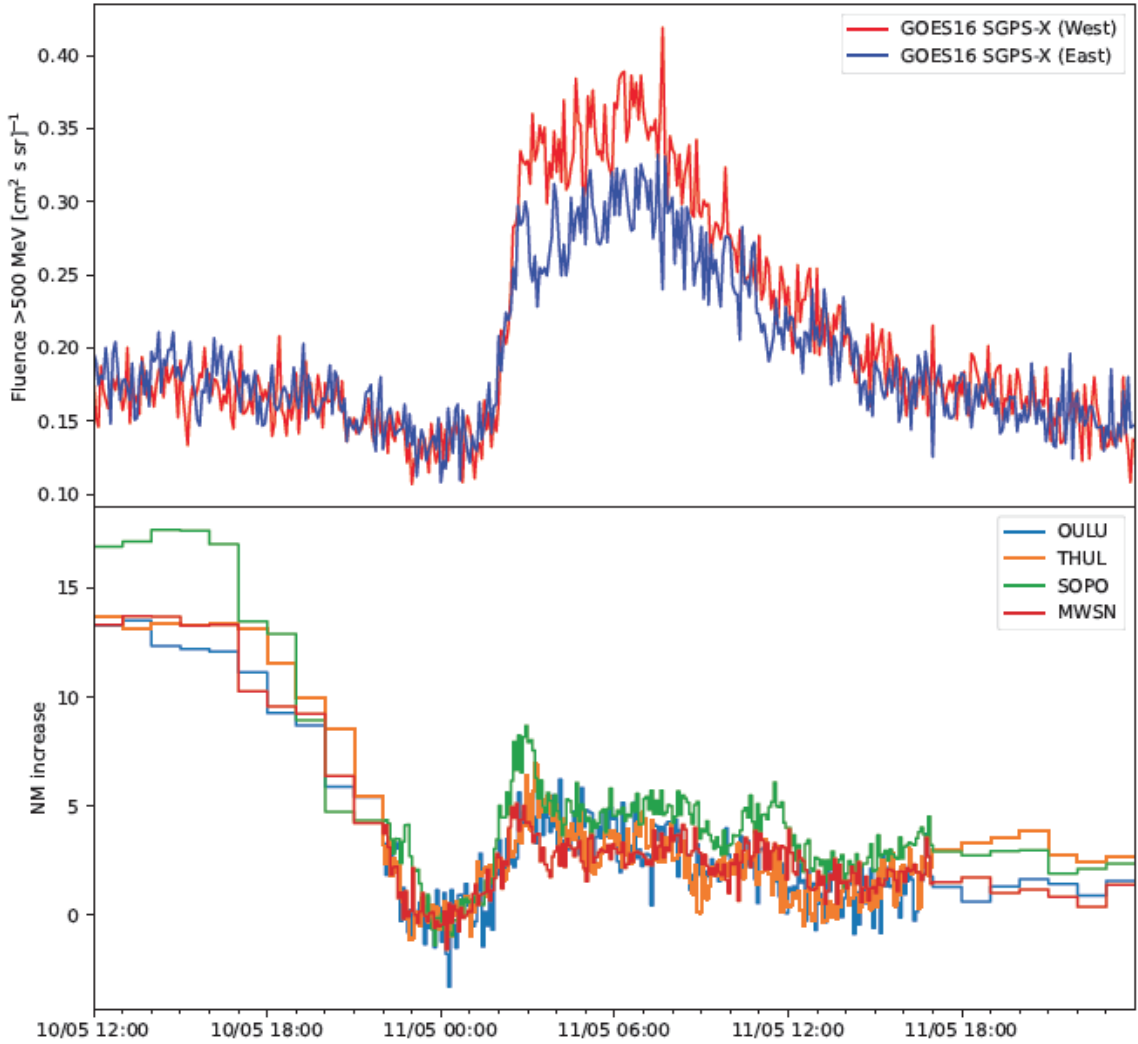

Figure 6: Cosmic Ray Flux >500 MeV observed by eastern- and western-facing monitors of the SGPS detector onboard the GOES-16 satellite, based on the NOAA data portal, alongside the increase in ground-based neutron monitor count rates from high-latitude regions. The neutron monitor sites include Oulu (located in Finland at N65°03′, E025°28′, 15 m above sea level), THUL (located in Thule, Greenland, at N76°30′, W068°42′, 26 m above sea level), SOPO (located at the South Pole in Antarctica, at S90°00′, N/A, 2820 m above sea level), and MWSN (located at Mawson in Antarctica, at S67°36, E062°53′, 30 m above sea level), as sourced from the Oulu Neutron Monitor and Neutron Monitor Database, and summarised in the International GLE Database. All data are corrected for barometric pressures without a detrending procedure. For neutron monitor increase calculations, the baseline was selected from 23:00 on 10 May to 01:00 on 11 May for MWSN, SOPO, and THUL, and from 23:30 on 10 May to 01:30 on 11 May for Oulu.

During the period under investigation in this study, specifically on 10 – 11 May 2024, we observed a significant variation in cosmic-ray activity. Corresponding to the arrival of CMEs at 17:05 UT, as shown by the SSC in the Kakioka Observatory (Kakioka Magnetic Observatory, 2013, see details in Section 6), a significant Forbush decrease (Forbush, 1937) in cosmic-ray intensities was observed in all available datasets with different cutoff rigidities. The maximum decrease in the neutron monitor count rate of magnitude $\geq 15\%$ or more was reached around midnight, after which the recovery phase began.

Around 2 UT on 11 May flux of high-energy (>500 MeV) protons started to increase as registered by the GOES-16 satellite (Figure 6). Notable variation of the solar proton flux triggered alerts e.g by the Bartol group, the Athens GLE alert++ by the Athens group, and an alert issued at 01:55 UT by the Kazakhstan group (private communication with O. Kraykunova). However, the two latter alerts were produced using several mid-latitude stations and are likely due to a reduction in the cutoff rigidity resulting from strong magnetospheric storms (Kudela *et al.*, 2008). Detailed neutron monitor records are provided in the Neutron Monitor Database (Mavromichalaki *et al.*, 2011) and International GLE database (Usoskin *et al.*, 2020). According to a careful inspection of the records, polar neutron monitors, such as the South Pole, Thule, Oulu, and Mawson (Figure 6), exhibit a small increase from $\approx 2$ UT to 10 UT during the recovery of the deep Forbush decrease, whose profile corresponds to the SEP profile registered by the SGPS onboard the GOES-16 satellite. Since the cutoff rigidity of these neutron monitor stations is nearly zero, the magnetospheric effect, that is, the reduction of the cutoff rigidity, cannot be used to explain the count rate increase.

Therefore, according to the recently updated GLE definition: *"A GLE event is registered when there are near-time coincident and statistically significant enhancements of the count rates of at least two differently located high-elevation neutron monitors and a corresponding enhancement in the proton flux measured by a space-borne instrument(s)"* (for details see Poluianov *et al.*, 2017, and the discussion therein), the global neutron monitor network observed a GLE. Therefore, this event was confirmed as GLE #74, the second of the current Solar Cycle 25, after GLE #73 on 28 October 2021 (Papaioannou *et al.*, 2022; Mishev *et al.*, 2022). Accordingly, this GLE was included in the International GLE Database. As discussed above, the neutron monitor count rate increases during this event are a complicated interplay between the recovery of the Forbush decrease, increased SEP flux, and geomagnetic storm impact; the former plays a role at polar stations and the latter at mid-and low-latitude stations. A detailed analysis of this complex event is desired and reserved for further dedicated studies.

**5. Interplanetary Shock Arrivals and Magnetospheric Compressions**

We used datasets of measurements conducted by the Time History of Events and Macroscale Interactions during Substorms (THEMIS) spacecraft (Angelopoulos, 2008) to evaluate the magnetospheric compressions upon shock arrivals. We derived the magnetic field measurements from fluxgate magnetometers (Auster *et al*., 2008) and ion moments from an electrostatic analyser (McFadden *et al*., 2008). When the interplanetary shock arrived on 10 May 2024, all three THEMIS satellites were aligned within the post-noon sector of the dayside magnetosphere during the inbound segment of their orbits.

Figure 7 shows the THEMIS-E measurements of the magnetic field and plasma from 17:04–17:07 UT on 10 May. From top to bottom, it shows the magnetic field strength ($B_{tot}$), magnetic field components ($B_x$, $B_y$, $B_z$) in the Geocentric Solar Ecliptic (GSE) coordinate system, ion-energy flux spectrogram, bulk-flow velocity ($V_x$, $V_y$, $V_z$) in the GSE coordinate system, and number density ($N_i$), respectively. The GSE spacecraft position is annotated in the lower part of the plot. Additionally, the vertical dashed line denotes the moment of the magnetopause crossing (17:05 UT) at x = 7.1 $R_E$, y = 3.2 $R_E$, and z = –2.7 $R_E$, indicating that THEMIS-E crossed over the magnetopause at ≈ 8.24 $R_E$. Figure 7 shows the sudden offsets of all magnetic field and plasma parameters, particularly the particle-energy dispersion, at the magnetopause. Prior to the magnetopause, we can observe a period of a stable magnetic field with a positive $B_z$ component accompanied by a low bulk-flow velocity and high concentration of high-energy ions (> 1 keV), which are typically observed in the Earth's magnetosphere. The region between the dashed and solid lines corresponds to the magnetosheath, which is characterised by a weaker magnetic field, broader ion-energy spectrum, and high ion-number density. The area after the solid line indicates the THEMIS-E transition into solar wind, characterised by a high plasma-flow velocity and narrow ion-energy spectrum of approximately 1 keV.

Figure 8 shows the THEMIS-A measurements of the magnetic field and plasma from 17:04–17:07 UT on 10 May. These datasets independently confirm the estimated values. The vertical dashed line denotes the moment of the magnetopause crossing (17:05 UT) at the [x, y, z] coordinates in the Earth's radius ($R_e$) of x = 7.7 $R_e$, y = 2.4 $R_e$, z = −2.7 $R_e$, indicating that THEMIS-A crosses over the magnetopause at ≈ 8.51 $R_e$, which is consistent with THEMIS-E's crossing at ≈ 8.24 $R_e$.

Figure 9 shows the THEMIS-E measurements of the magnetic field and plasma data in the same order as that in Figure 7. In this case, the spacecraft crosses from the magnetosheath to the magnetosphere, encountering the magnetopause at x = 3.4 $R_e$, y = 3.2 $R_e$, and z= − 1.9 $R_e$ at approximately 19:12 UT, as denoted by the dashed line. This indicates that THEMIS-E traverses the magnetopause at ≈ 5.04 $R_e$. Initially, the magnetic field exhibits an intense negative $B_z$ component, which subsequently undergoes significant rotation within the magnetosheath region. The ion bulk-flow velocity is also high and fluctuates substantially in this region. A wider ion-energy spectrum is discernible here. Upon entering the magnetospheric region, the magnetic field was still strong but devoid of high-frequency disturbances. Additionally, as expected, the ion bulk-flow velocity of magnetospheric plasma was considerably lower; however, the number density was higher than normal.

This compressed magnetopause value (≈ 5.04 $R_e$) is somewhat extreme and lower than that of the geosynchronous orbit (6.6 $R_e$). Figure 10 shows the THEMIS-A measurements of the magnetic field and plasma from 18:50–19:30 UT on 10 May. The vertical dashed line denotes the moment of magnetopause crossing (19:19 UT) at x = 4.0 $R_e$, y = 2.9 $R_e$, and z = −2.1 $R_e$, indicating that THEMIS-A crossed over the magnetopause at ≈ 5.37 $R_e$. This is broadly consistent with THEMIS-E's crossing at ≈ 5.04 $R_e$. Here, we considered entry into the magnetosphere to occur when the energy of the main particle population exceeds 1 keV and particles from the magnetosheath are absent, which are typical characteristics of the magnetosphere. The period preceding the dashed lines represents particles with a wide range of energies, indicative of a mixture of the magnetosphere and magnetosheath plasma. This region can be identified as a boundary layer characterised by open magnetic field lines. Overall, these measurements confirmed that the magnetosphere compressed considerably immediately after the shock arrival at 17:05 UT. However, further assessments of this extreme magnetospheric compression are required.

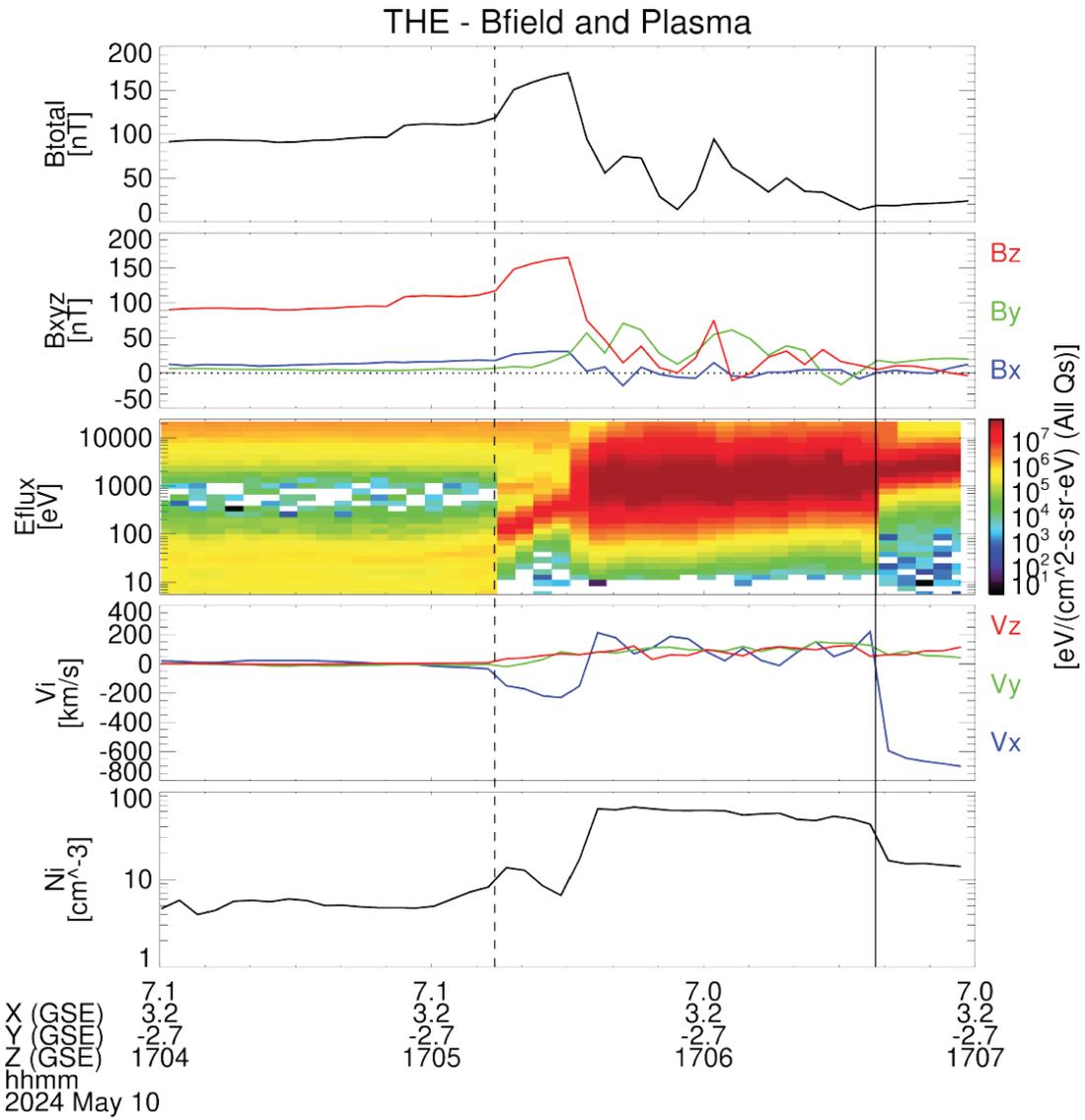

Figure 7: THEMIS-E's measurements of magnetic field strength ($B_{total}$), magnetic field components ($B_x$, $B_y$, $B_z$) in the GSE coordinate system, ion-energy flux spectrogram, bulk-flow velocity ($V_x$, $V_y$, $V_z$) in the GSE, and number density ($N_i$) from 17:04–17:07 UT on 10 May. The magnetopause crossing is indicated by a vertical dashed line.

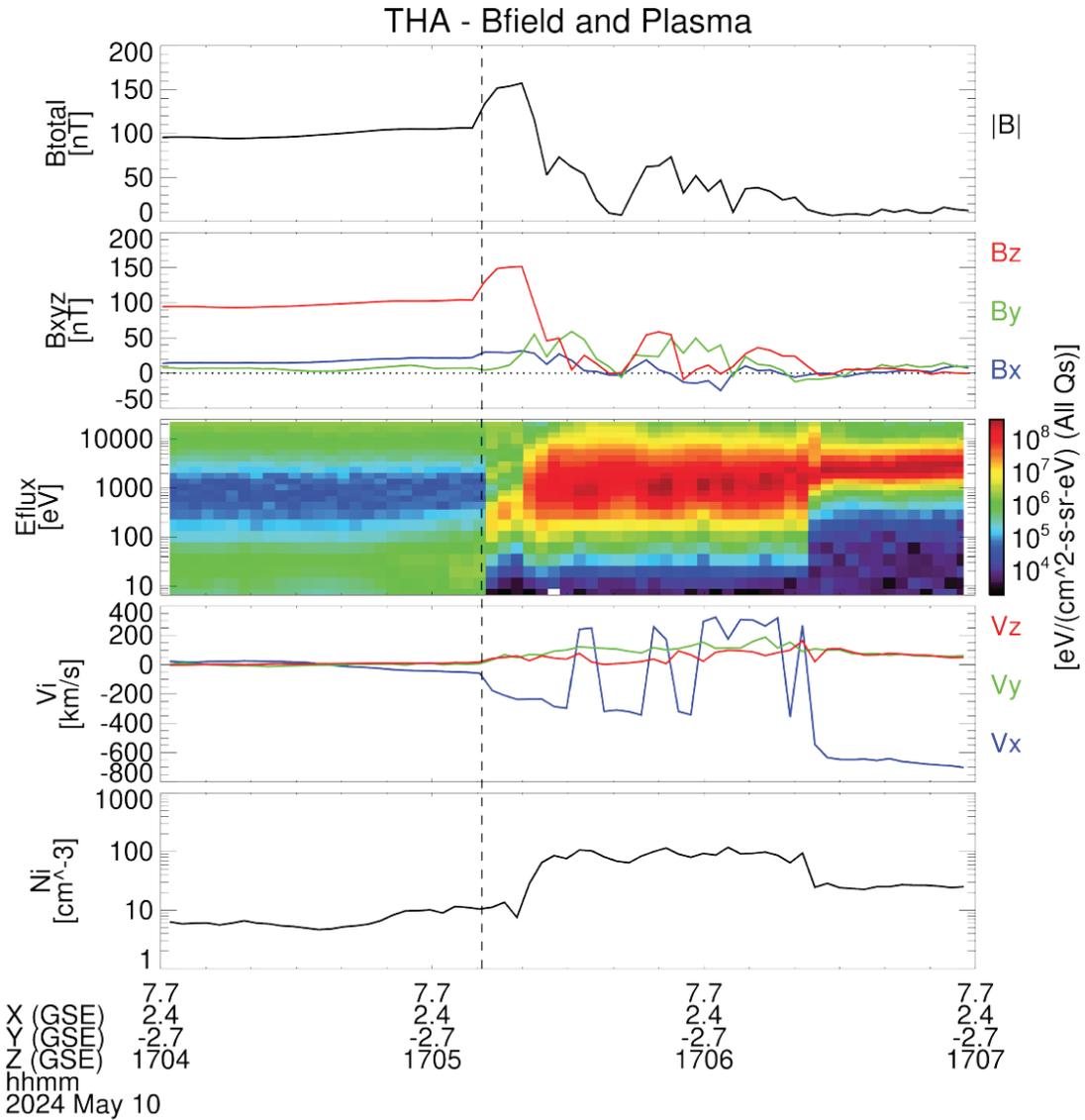

Figure 8: THEMIS-A's measurements of magnetic field strength ($B_{total}$), magnetic field components ($B_x$, $B_y$, $B_z$) in the Geocentric Solar Ecliptic (GSE) coordinate system, ion-energy flux spectrogram, bulk-flow velocity ($V_x$, $V_y$, $V_z$) in GSE, and number density ($N_i$) from 17:04–17:07 UT on 10 May. The magnetopause crossing is indicated by a vertical dashed line.

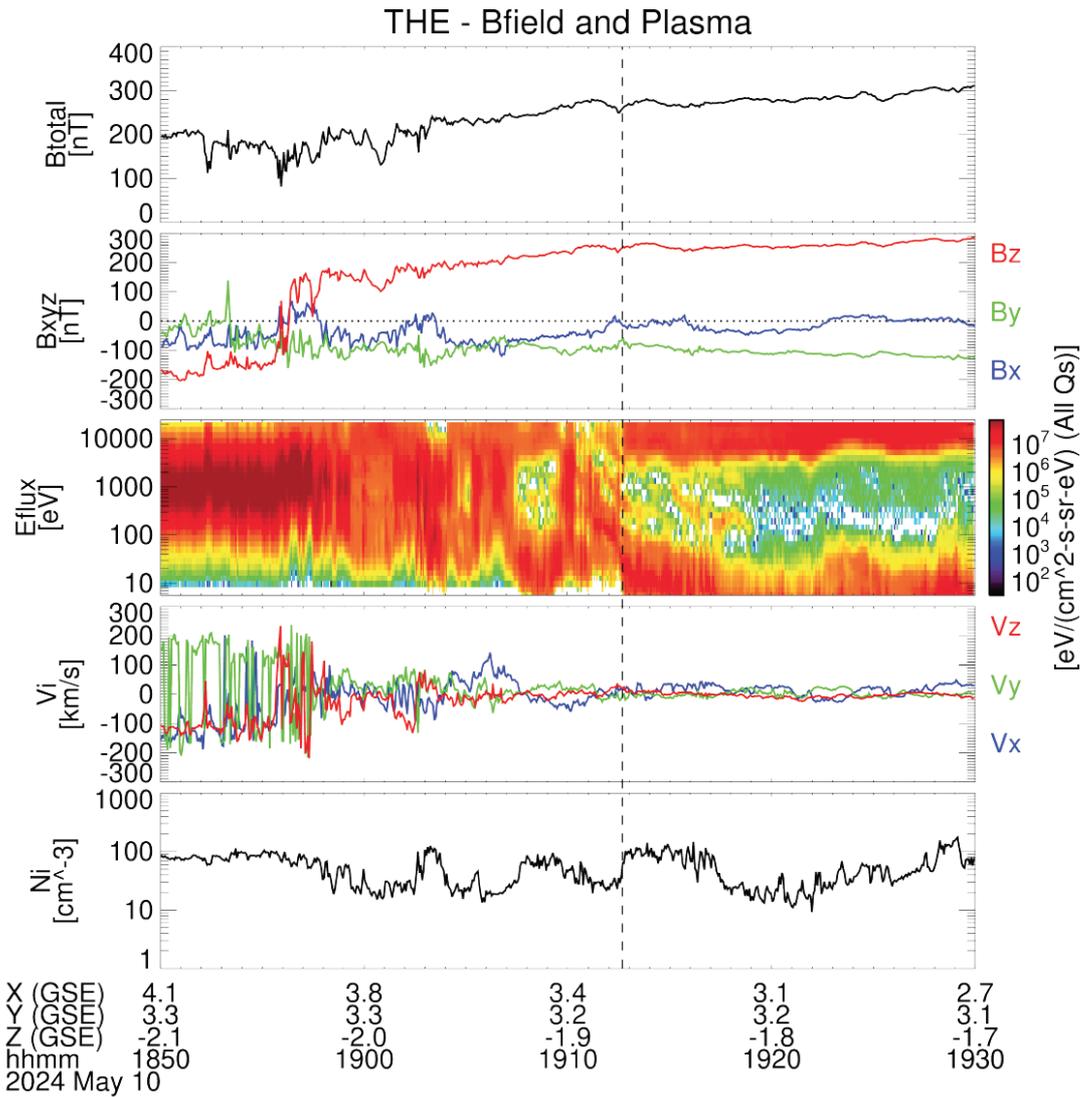

Figure 9: THEMIS-E's measurements of magnetic field strength ($B_{total}$), magnetic field components ($B_x$, $B_y$, $B_z$) in the Geocentric Solar Ecliptic (GSE) coordinate system, ion-energy flux spectrogram, bulk-flow velocity ($V_x$, $V_y$, $V_z$) in GSE, and number density ($N_i$) from 18:50–19:30 UT on 10 May. The magnetopause crossing is indicated by a vertical dashed line.

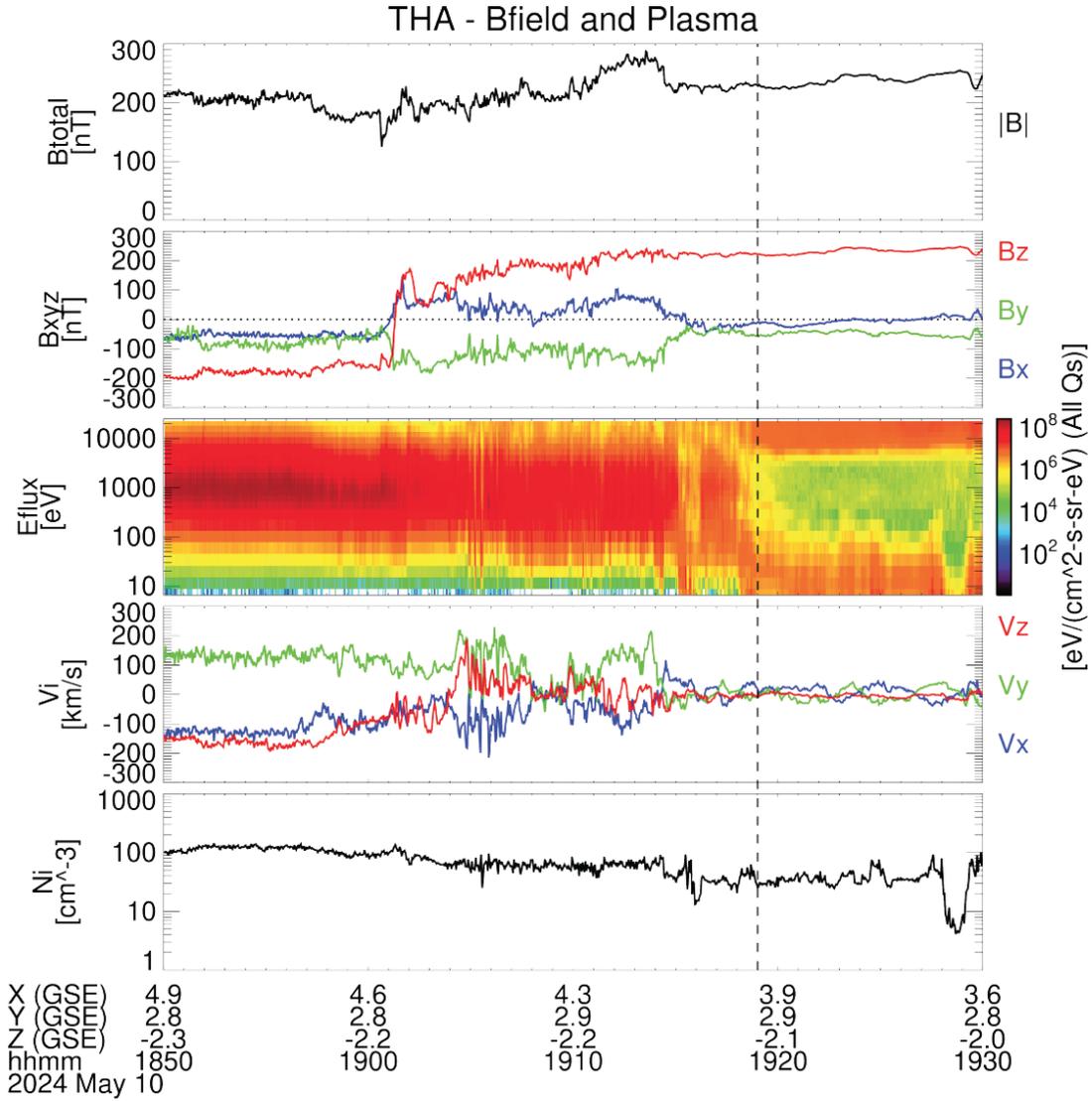

Figure 10: THEMIS-A's measurements of magnetic field strength ($B_{total}$), magnetic field components ($B_x$, $B_y$, $B_z$) in the Geocentric Solar Ecliptic (GSE) coordinate system, ion-energy flux spectrogram, bulk-flow velocity ($V_x$, $V_y$, $V_z$) in GSE, and number density ($N_i$) from 18:50–19:30 UT on 10 May. The magnetopause crossing is indicated by a vertical dashed line.

## 6. Geomagnetic Disturbances

The geomagnetic measurements captured at least four SSCs in May 2024. The Kakioka $\Delta H$ measurement recorded SSCs at $\approx$ 14:07 UT on 2 May (26 nT), 17:05 UT on 10 May (78 nT), 05:51 UT on 16 May (10 nT), and 13:26 UT on 17 (18 nT), respectively (Kakioka Magnetic Observatory, 2015). According to the real-time solar wind data acquired by ACE, a jump in solar wind speed and density arrived at the L1 Lagrange point at 16:36 UT on 10 May. In the vicinity of the jump, the speed increased from approximately 460 to 660 km/s, and the density increased from $\approx$ 10 to 27 cm$^{-3}$. This jump caused a substantial SSC from $\approx$ 17:05 UT on 10 May up to $\approx$ 78 nT in the Kakioka $\Delta H$ measurement (Kakioka Magnetic Observatory, 2015). This implies that the travel time from the L1 point to Earth is 29 min at the SSC moment. This estimate includes the transition time from the magnetopause to the ground.

Figure 11 summarises $\Delta X$ observed at 20° < MLAT (magnetic latitude) < 30° around the SSC. Again, $\Delta X$ at 17:04 UT on 10 May 2024 (immediately before the SSC) is set to be zero as a baseline. Abrupt increase in $\Delta X$ is evident at 17:05 − 17:06 UT. The SSC amplitude of the abrupt increase in $\Delta X$ is larger on the duskside than on the dawnside. This is consistent with the previous studies and attributed to the contribution from the magnetospheric current and the ionospheric current as well (*e.g.*, Araki *et al.*, 2006). The European sector witnessed a relatively large SSC amplitude, as it was located in the dusk side in the summer hemisphere upon the SSC. This favourable location probably enhanced ionospheric conductivity and the field-aligned current to increase the magnetic variations in the European sector (Shinbori *et al.*, 2009, 2012). The amplitude is $\approx$ 130 nT, which is large but not unique in comparison with the historical records of SSC (Araki, 2014; Hayakawa *et al.*, 2022).

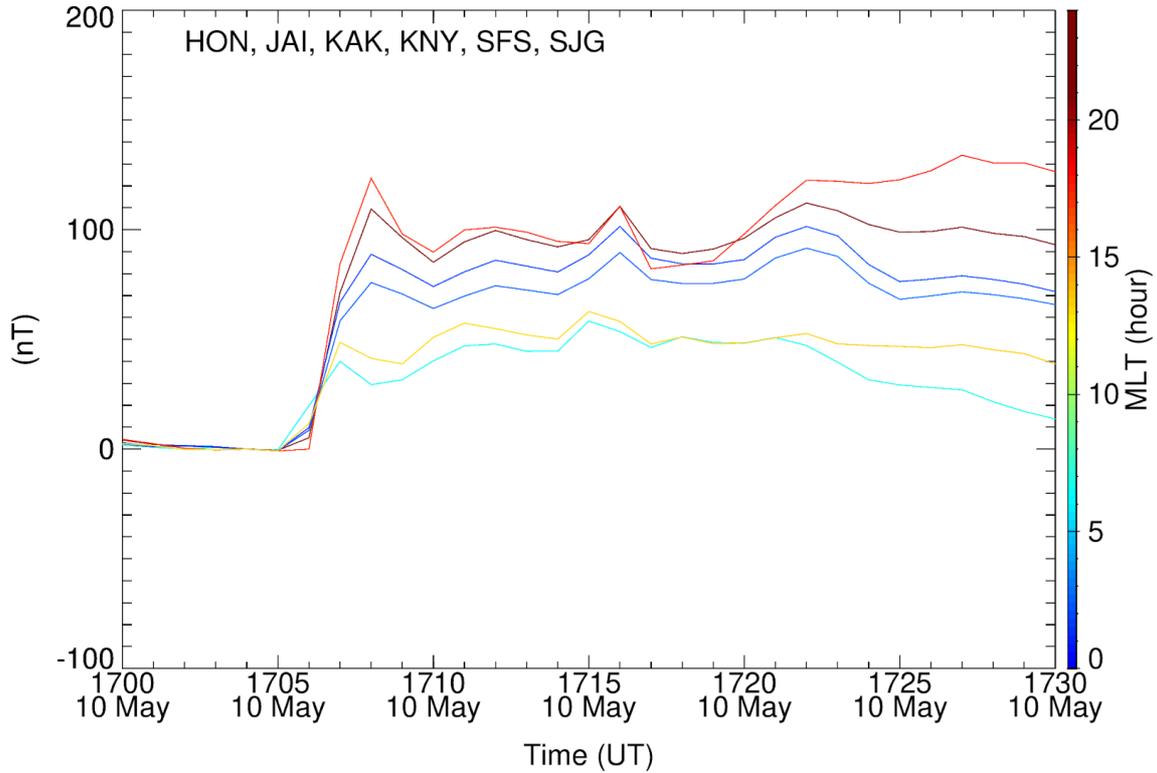

Figure 11: The north component (ΔX) of the geomagnetic disturbances at 20° < MLAT < 30°. The colour codes indicate MLT (magnetic local time). The abbreviation codes indicate the sites used to plot.

Figure 12 summarises measurements of IMF and solar wind parameters, the SME index, the AE and AL indices, the Dst and Dxt global indices, the Dxt local indices, and the absolute values of the magnetic latitudes at which the aurorae and stable red auroral (SAR) arcs were reported. The IMF and solar wind parameters are acquired from the 1-min OMNI data (Papitashvili and King, 2020). We note that the IMF and the solar wind parameters were observed by the ACE spacecraft at 245 $R_k$ upstream of the solar wind, and the data were time-shifted to the subsolar point of Earth's bow shock. Thus, the time-variations are not always correct due to ambiguity of the time-shifting algorithm, in particular, during the highly disturbed conditions. Because of this reason, we intend to overview the IMF and solar wind parameters briefly, without going into detail, although the IMF datasets probably indicate at least one compressed flux rope inside the sheath of the subsequent CME(s) and the density profile indicates arrivals of multiple sheaths.

The SSC (Figure 11) was caused by rapid increases in the solar wind velocity from ≈ 450 km/s to > 700 km/s and the solar wind density from ≈17 cm³ to >30 cm³. At the same time, the magnitude of the auroral electrojet increased as indicated by the AE, AL, SMU and SML indices. The IMF $B_z$ turned southward just after the SSC, reaching $\approx -40$ nT around 18:08 UT. After the SSC, there are at least 9 intervals in which IMF $B_z$ remained negative for 30 minutes and more. They are summarised in Table 3. The solar wind speed gradually increased to 1026 km/s at $\approx 01:17$ UT on 12 May.

Table 3: Intervals in which IMF $B_z$ remained negative for 30 minutes and more. IMF $B_z$ is obtained by the 1-min OMNI data.

|   | Start Time | End Time | Duration (minutes) | Minimum IMF $B_z$ (nT) |
|---|---|---|---|---|
| 1 | 19:41 UT on 10 May | 22:23 UT on 10 May | 162 | $-40$ |
| 2 | 22:41 UT on 10 May | 01:36 UT on 11 May | 175 | $-56$ |
| 3 | 01:45 UT on 11 May | 04:23 UT on 11 May | 158 | $-35$ |
| 4 | 06:07 UT on 11 May | 07:37 UT on 11 May | 90 | $-40$ |
| 5 | 07:57 UT on 11 May | 11:01 UT on 11 May | 184 | $-41$ |
| 6 | 11:29 UT on 11 May | 17:39 UT on 11 May | 360 | $-25$ |
| 7 | 20:18 UT on 11 May | 20:48 UT on 11 May | 30 | $-9$ |
| 8 | 23:06 UT on 11 May | 23:49 UT on 11 May | 43 | $-7$ |
| 9 | 00:12 UT on 12 May | 02:48 UT on 12 May | 156 | $-5$ |

The Dst/Dxt indices rapidly decrease during the sustained southward IMF intervals of 1 and 2 listed in Table 3. They reached a minimum value of $-412$ nT / $-415.5$ nT, respectively during the sustained southward IMF interval 3. This marks the end of the storm main phase and the storm peak hour, and indicates a considerably fast decrease in Dst/Dxt with peak-to-peak amplitudes of 474 nT and 479.8 nT in 9 hours. According to Tsubouchi and Omura (2007), it is possible to calculate a statistical return period of geomagnetic storms of this minimum value (min Dst = $-412$ nT) as 5.7 years. The real-time Dst/Dxt values are subject to small changes until definite values are established (WDC for Geomagnetism at Kyoto *et al.*, 2015; Karinen and Mursula, 2005; Mursula *et al.*, 2023).

Figure 12 shows that the Dst and Dxt indices follow each other quite closely, with momentary differences varying between -9nT and +6nT. Over the depicted time interval, the Dxt index is only slightly (less than 1 nT) lower than the Dst index. These differences are due to the somewhat different normalizations of the two indices (Mursula *et al.*, 2008). The multiple complicated structures of ICMEs (Section 3) likely caused the two-step development of this storm. The intensification of the ring current was caused by the penetration of hot ions into the deep inner magnetosphere due to the dawn-dusk convection electric field intensified by the large amplitude southward IMF and fast solar wind. During the main phase of the storm, the solar wind speed was as high as ≈ 700–800 km/s, and the southward component of the IMF was as high as ≈ 50 nT. The solar wind speed, maximum value of the southward component of the IMF, and minimum Dst index were similar to those of the large magnetic storm of 20–21 November 2003. The hot ions are speculated to penetrate as low as $L ≈ 1.5$ to largely intensify the ring current (Ebihara *et al.*, 2005).

Figure 12 also shows the local Dxt indices at the four Dst/Dxt stations. Their average makes the Dxt index. The largest compression at 17 UT on 10 May, close to the SSC occurrence, is observed at SJG (+93.9 nT) which is located at this time closest of all stations to noon, in the early afternoon of local time. The main phase development proceeds fairly similarly in all the four stations (at different local times) until 22 UT on 10 May, after which somewhat larger differences start appearing. Note that only one station, HER, reaches its local Dxt minimum at 2 UT on 11 May, at the time of the (global) Dst/Dxt minimum, while SJG has it one hour before, HON two hours afterward at 4 UT and KAK only at 8 UT, six hours after the Dxt minimum.

After 2 UT on 11 May, the Dst/Dxt indices started increasing, and the storm recovery phase began. During the recovery, the Dst/Dxt indices show some negative excursions corresponding to the southward turnings of the IMF $B_z$. Note that the local time of both KAK and HON at the time of their minimum observation is about 17 LT. For SJG it is 20 LT and for HER 3 LT. This is in agreement with the fact that, statistically, at low latitudes, the largest disturbance is found around 18 LT (Cummings, 1966; Yakovchouk *et al.*, 2012). This local-time dependence of the storm-time geomagnetic disturbances is often thought to be the partial ring current, but is probably not simple (Fukushima and Kamide, 1973). According to in-situ observations, the peak of the ring current takes place in the dusk-midnight sector (Terada *et al.*, 1998; Ebihara *et al.*, 2002; Le *et al.*, 2004). This is different from the ground-based magnetic observations. This discrepancy is probably attributed to the contribution from the field-aligned current (Ohtani *et al.*, 2007).

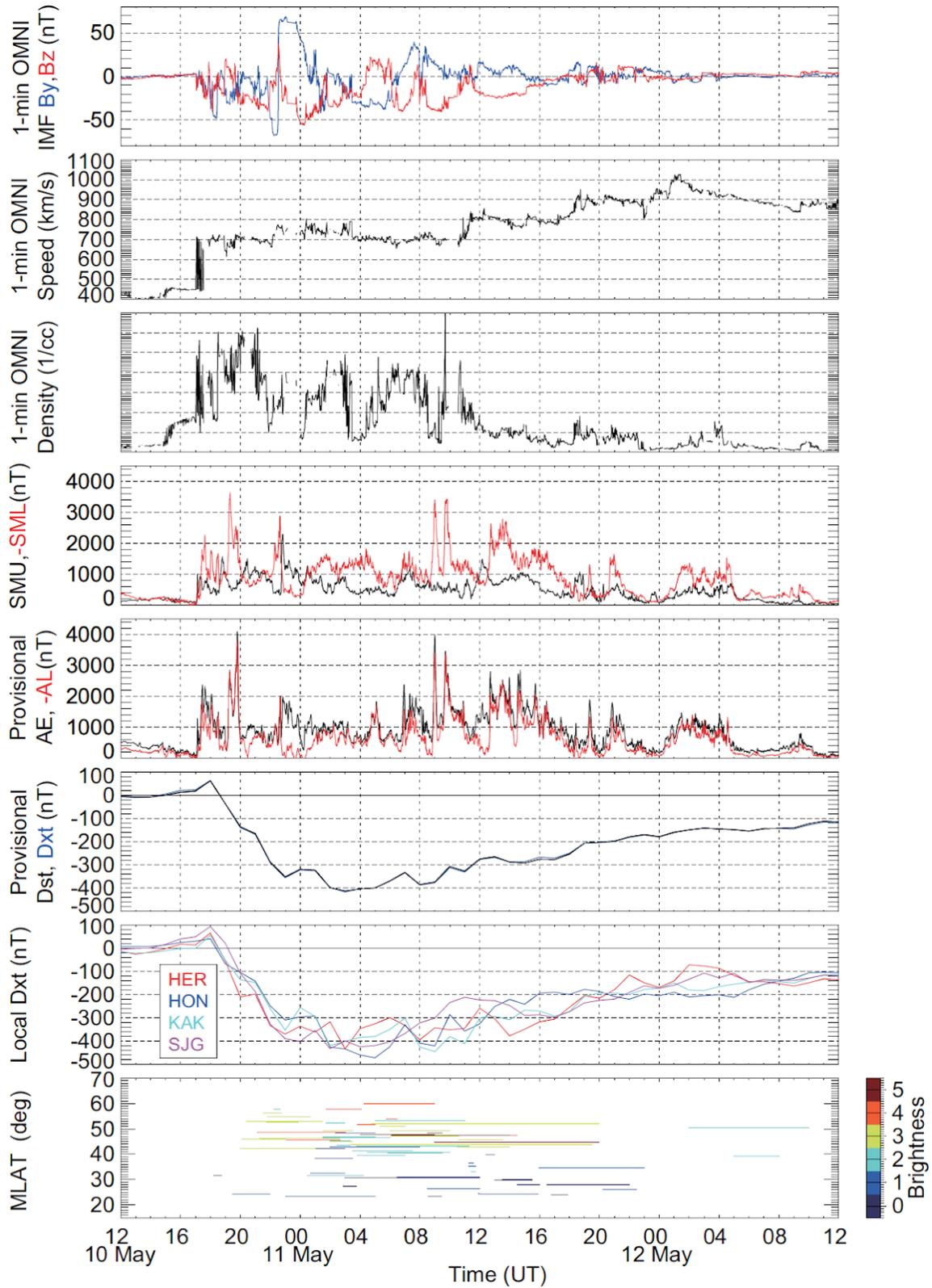

Figure 12: Temporal evolutions of interplanetary magnetic field (IMF), solar wind speed and density, geomagnetic disturbance, and auroral visibility. The first panel shows the interplanetary magnetic

field of the solar wind in $B_x$ (blue) and $B_z$ (red) in the GSM coordinates. The second panel shows the solar wind velocity. The third panel shows the density of the solar wind. For these three panels, we have derived their source data from the 5-min OMNI data. The IMF and the solar wind data are time-shifted to the subsolar point of Earth's bow shock. Thus, the variations are subject to changes depending on the calculation algorithm. The fourth panel shows the SME (black) and SML (red) indices. The fifth panel shows provisional AE (black) and AL (red) indices. The sixth panel indicates the (global) Dst and Dxt indices and the seventh panel the local Dxt indices at Hermanus (HER), Honolulu (HON), Kakioka (KAK), and San Juan (SJG). The last panel displays the absolute value of the magnetic latitude at which the aurorae and SAR arcs were reported both by naked eyes and instruments. The brightness is generally given with International Brightness Coefficient (IBC classification), apart from 0 for brightness not enough for naked-eye visibility and 5 for brightness more than full moon.

As shown in Figure 12, the auroral electrojet has also developed significantly. The provisional AE and AL indices (WDC for Geomagnetism at Kyoto *et al.*, 2015b) initiated significant development at 17:06 UT on 10 May and increased to 2379 nT and −1830 nT, respectively, at 17:28 UT. These peak values were reached in 22 min. The abrupt jump in the provisional AE index is probably regarded as a shock-triggered substorm (Akasofu and Chao, 1980; Tsurutani and Zhou, 2003). The ionospheric current associated with the main impulse of the SSC can also contribute to the abrupt jumps in these indices (Araki, 1994; Zhang *et al.*, 2023). During the storm's main phase, the provisional AE and AL indices reached peak values of 4098 and −3797 nT, respectively, at 19:48 UT on 10 May. Subsequently, these indices remained within < |2000| nT during the main storm phase. During the storm recovery phase, these indices reached peak values of 3982 and −3797 nT at 09:00 UT on 11 May. According to a statistical study by Nakamura *et al.* (2015), up to 16 events occurred during which AL exceeded −3000 nT from 1996 to 2012, and the recurrence period for AL of −3797 nT was 1.49 years. In this sense, the provisional AL index of −3797 nT was not that extreme. We have to note that the AE and AL indices are not necessarily proportional to the intensity of the auroral electrojet.

Based on the behaviours of the AE index and the AL index around the storm peak, the auroral oval is suspected to extend significantly equatorward from the AE stations during this interval, resulting in the suppression or saturation of the AE and AL indices. The SME index is used to overcome this issue since it is calculated from many sites, usually exceeding 100 (Newell and Gjerloev, 2011). In that sense, the SME index is better. However, the AE index has been used to represent the magnitude of the auroral electrojet. We believe that AE is still useful to compare with historical events, and to drive empirical models depending on AE. In order to see this possibility, we have also checked the SME including the SML index in Figure 12. The fourth panel of Figure 12 shows the SME index. Interestingly, the peak amplitude of the SML index, which is an extension of the AL index, is not always larger than that of the AL index. These values are provisional, so that the definitive conclusion cannot be obtained regarding the auroral electrojet.

Figure 13 summarises the north component of the geomagnetic disturbances $\Delta X$ at different latitudinal zones. The data was provided by INTERMAGNET (2021). The sites used to plot are summarised in Table 4. $\Delta X$ at 17:04 UT on 10 May 2024 is set to be zero. Diurnal variations, such as Sq, were not removed. Figure 13a shows $\Delta X$ at MLAT > 75°. After the SSC, $\Delta X$ is negative on the duskside, whereas it is positive on the dawnside. The variations are reasonably explained by the twin vortices of the Hall current flowing in the ionosphere, known as the DP2 equivalent current (Nishida, 1968). In the poleward part of the DP2 current, it tends to flow westwardly on the duskside, and eastwardly on the dawnside as schematically illustrated at the right-bottom corner of Figure 13. Figure 13b shows $\Delta X$ in the latitudinal zone, 60° < MLAT < 70°, which is usually referred to as the auroral zone. The lower envelope of $\Delta X$ in this panel is roughly equivalent to the AL index. The westward current is the most intense around midnight and in the postmidnight, which is consistent with the previous studies (*e.g.*, Kamide, 1982). In general, $\Delta X$ is positive on the duskside, and negative on the dawnside. This is consistent with the expectation that $\Delta X$ is primarily, not all, caused by the equatorward part of the DP2 current. Figure 14c shows $\Delta X$ in the lower latitudinal zone, 20° < MLAT < 30°. The contribution from the ionospheric current is likely to be small. The contributions from the ring current, the tail current, the magnetopause current and the field-aligned current are probably significant. The averaged $\Delta X$ should be closed to the Dst index. During the main phase, the variations of $\Delta X$ show almost coherently with respect to MLT as mentioned above, but it is clearly shown that $\Delta X$ tends to be smaller on the duskside than on the dawnside. After the storm peak, the MLT dependence became significant. There are some negative peaks at $\approx$ 02 UT, $\approx$ 08 UT, and $\approx$ 14 UT on 11 May. These peaks took place in the premidnight sector, suggesting that they were primarily caused by the storm-time partial ring current. During the interval from $\approx$ 06 UT to $\approx$ 10 UT on 11 May, $\Delta X$ shows a negative excursion on the pre-midnight sector, whereas it shows a positive excursion in the pre-noon sector. This can be reasonably explained using the development of the partial ring current in the pre-midnight sector due to the injection of fresh ions from the nightside plasmasheet, and the reduction of the ring current in the pre-noon sector due to the escape of the pre-existing ions in response to the rapid enhancement of the magnetospheric convection (Hashimoto *et al*., 2002; Brandt *et al*., 2002; Ebihara *et al*., 2002).

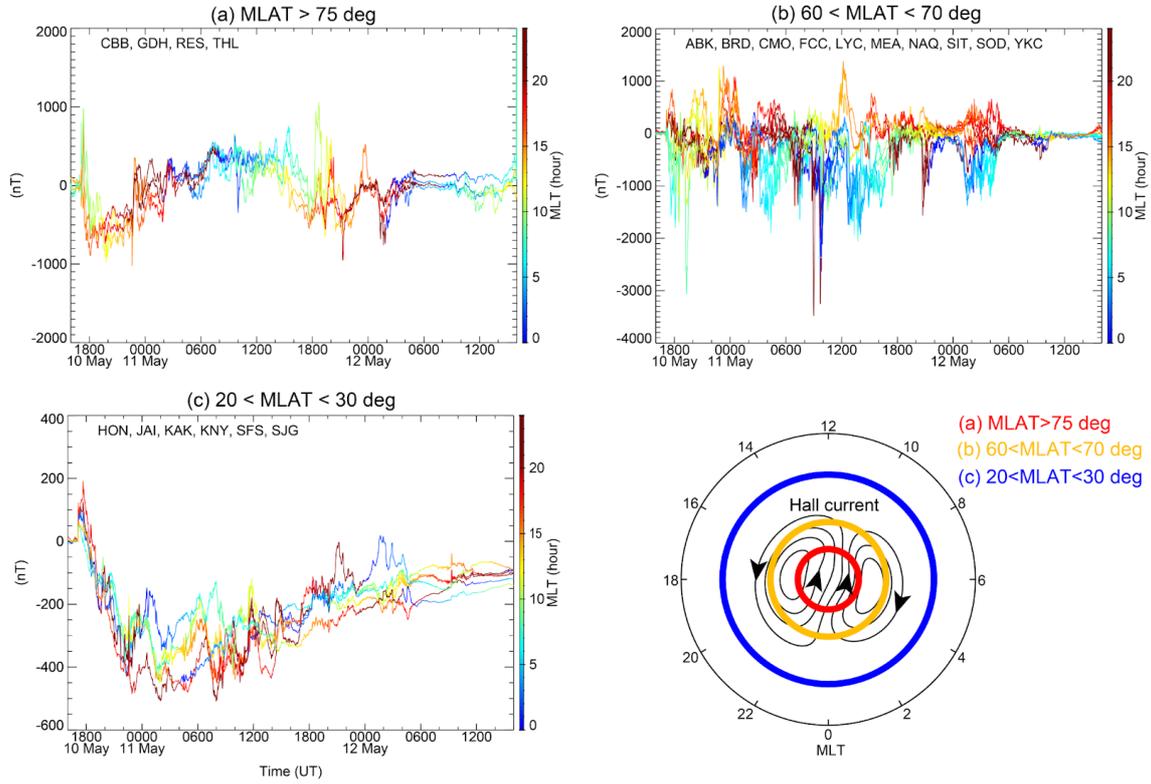

Figure 13: The north component of the geomagnetic disturbances ΔX at (a) MLAT > 75°, (b) 60° < MLAT < 70° and (c) 20° < MLAT < 30°, where MLAT stands for the magnetic latitude. The colour codes indicate MLT. The abbreviation codes indicate the sites used to plot.

Table 4: Geomagnetic observatory sites used to plot Figure 13.

| Abbreviation Code | Geographic Latitude (°) | Geographic Longitude (deg) |
|---|---|---|
| (a) MLAT > 75° | | |
| CBB | 69.12 | 254.97 |
| GDH | 69.25 | 306.47 |
| RES | 74.69 | 265.11 |
| THL | 77.47 | 290.77 |

| (b) 60° < MLAT < 70° | | |
|---|---|---|
| ABK | 68.4 | 18.8 |
| BRD | 49.87 | 260.03 |
| CMO | 64.87 | 212.14 |
| FCC | 58.76 | 265.91 |
| LYC | 64.6 | 18.7 |
| MEA | 54.62 | 246.65 |
| NAQ | 61.16 | 314.56 |
| SIT | 57.06 | 224.68 |
| SOD | 67.4 | 26.6 |
| YKC | 62.48 | 245.52 |
| (c) 20° < MLAT < 30° | | |
| HON | 21.32 | 202.0 |
| JAI | 26.92 | 75.8 |
| KAK | 36.23 | 140.19 |
| KNY | 31.42 | 130.88 |
| SFS | 36.7 | 354.1 |
| SJG | 18.11 | 293.85 |

## 7. Spatial Extents of the Auroral Visibility and the Auroral Oval: Naked Eye Observations

This storm significantly extended the auroral oval equatorward, as expected from the empirical correlation between the extension of the equatorward boundary of the auroral oval and the intensity of the associated geomagnetic storm (Yokoyama *et al*., 1998; Blake *et al*., 2020). As such auroral visibility was reported worldwide. We followed a citizen science approach to gather auroral visibility reports. This is a valid approach for reconstructing the equatorward boundary of auroral visibility and auroral oval, as reported by Case *et al*. (2016a). The authors prepared an online survey and called for auroral reports for the naked eye and instrumental visibility. This online inquiry allowed us to obtain 80 valid auroral reports, as exemplified with Figure 14. Figure 15 summarises responses from online survey responses. Our result is roughly consistent with independent survey datasets for auroral visibility during this storm (*e.g.*, Grandin and ARCTICS Collaboration 2024; Wicklund, 2024) with further possible equatorward extensions of the instrumental detections of the auroral emissions. We located their geographic coordinates and computed their magnetic latitudes using the coordinates of the northern geomagnetic pole (N80.8° W72.6°) in the IGRF-13 model (Alken *et al.*, 2021).

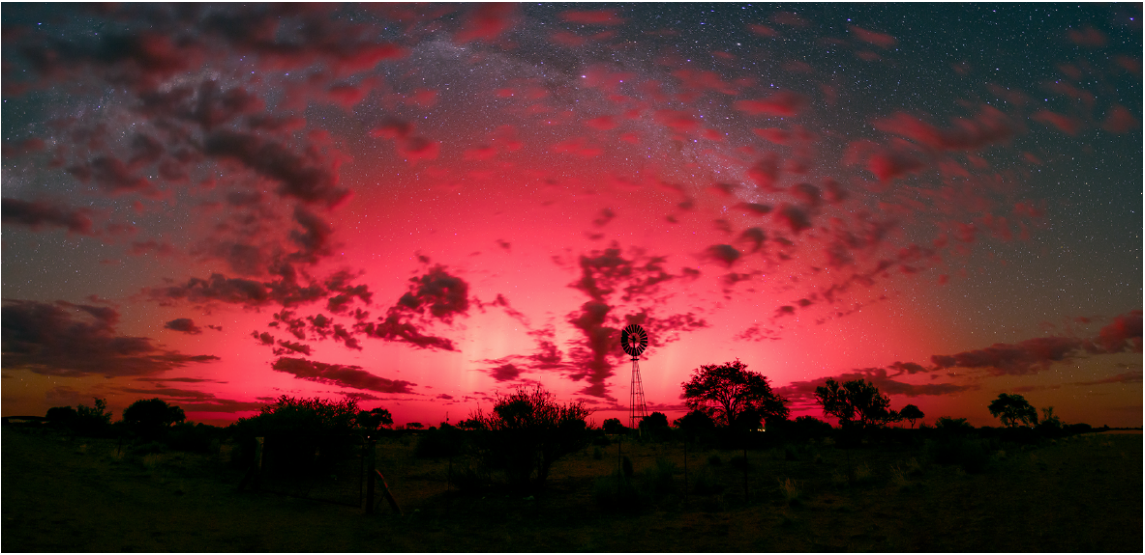

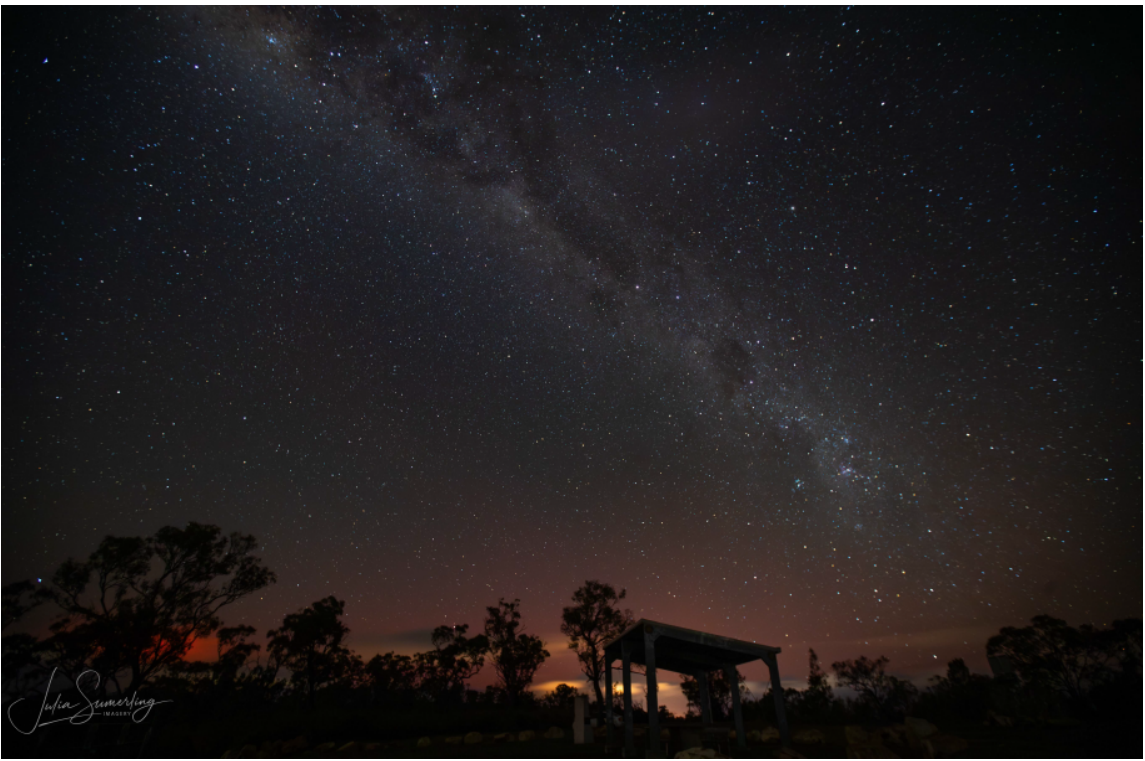

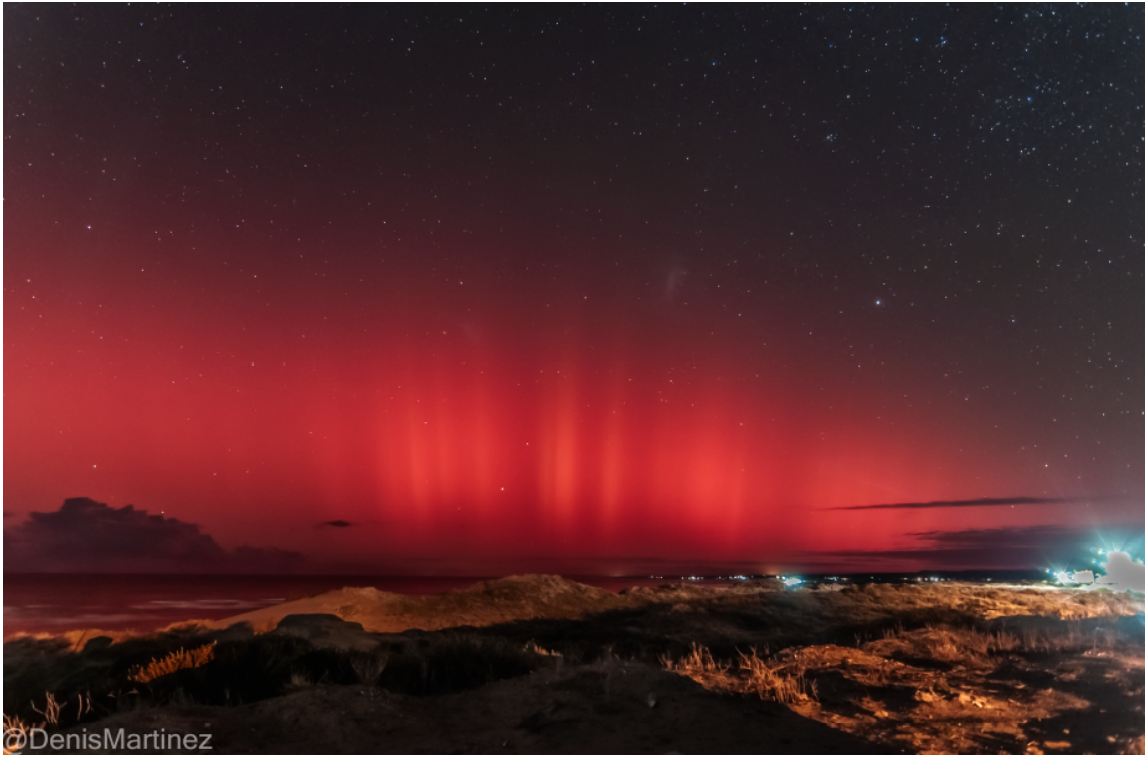

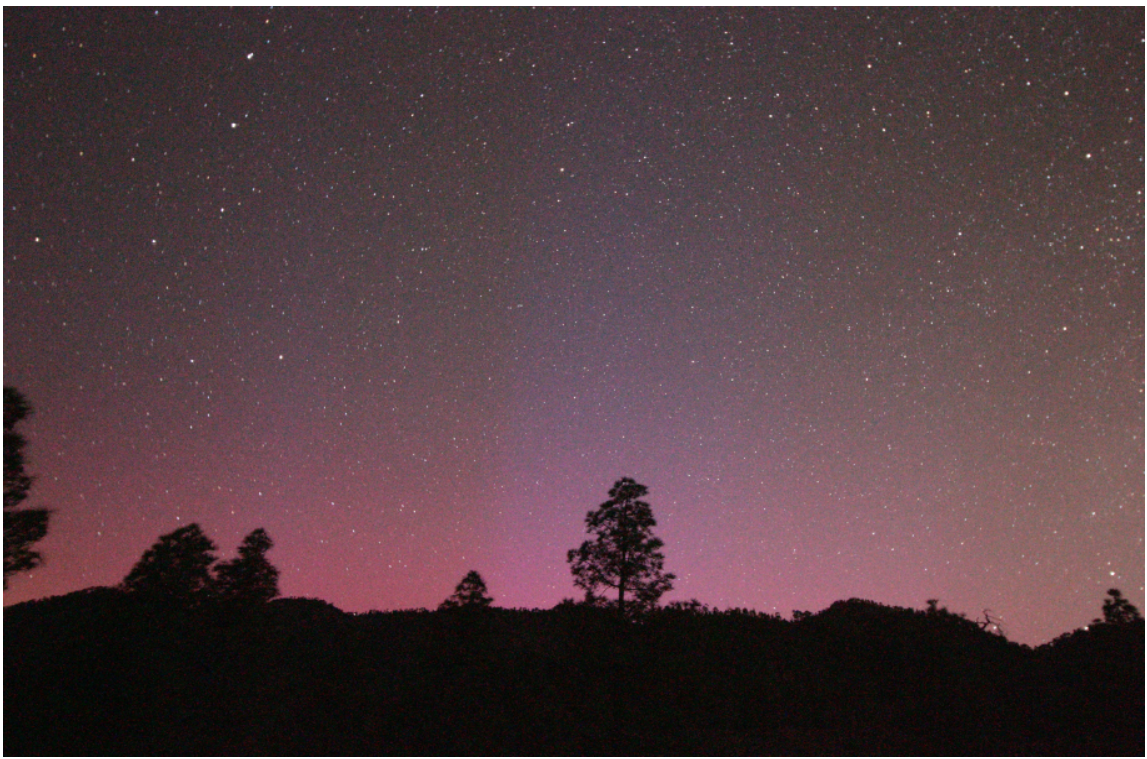

Figure 14: Examples of the reported auroral visibility throughout the world, from the top to the bottom: (a) Tivoli Astrofarm in Namibia (© Richard Payne), (b) Chewko Lookout near Mareeba in Australia (© Julia Sumerling),  (c) Las Grutas in Argentina (© Denis Martínez), and (d) Mogán in Gran Canaria (© Víctor R. Ruiz).

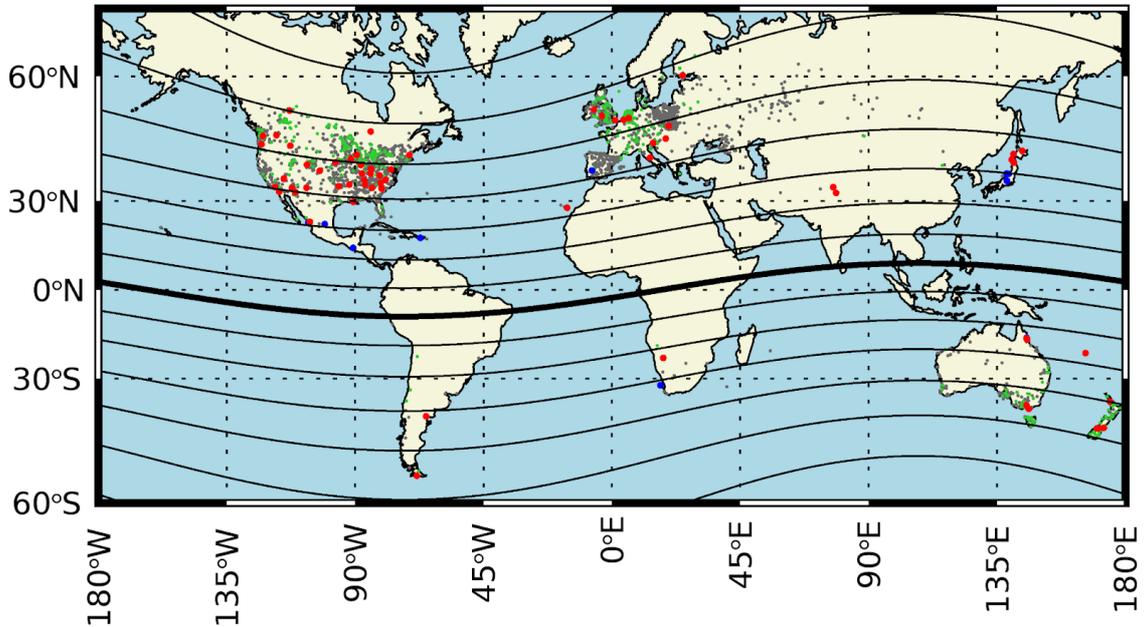

Figure 15: Geographical extents of the reported auroral visibility from 10 May 2024 to 12 May 2024. The contour lines indicate the magnetic latitude (MLAT) with an interval of each 10°. The solid line indicates the magnetic equator (0° MLAT). The red and blue dots indicated auroral report with naked-eye visibility (this section) and camera-visibility only (Appendix 1), respectively. For reference, we have plotted visibility reports from Grandin and ARCTICS Collaboration (2024) and Wicklund (2024) as greenish and grey small dots, respectively.

On this basis, we confirmed the equatorward extent of naked-eye auroral visibility in both hemispheres. In the Southern Hemisphere, naked-eye auroral reports have been confirmed down to Boulouparis in New Caledonia (S21°54′, E165°59′, −26.5° MLAT), Chewko Lookout in Australia (S17°03′, E145°22′, −24.2° MLAT), and Tivoli Astrofarm in Namibia (S23°28′, E018°01′, -23.2° MLAT). In the Northern Hemisphere, naked-eye auroral reports have been confirmed down to El Peyote (N22°41′, W100°36′, 30.7° MLAT) and Hanle in India (N32°28′, E078°34′, 24.3° MLAT).

Auroral visibility at any site does not immediately indicate overhead auroral visibility. Therefore, we must reconstruct the equatorward boundaries of the auroral oval considering the MLAT and the reported altitude at each site. The altitude profiles were used to reconstruct the equatorward boundaries of the auroral oval conservatively. We assumed the altitude of the visual aurorae to be approximately 400 km, following Roach *et al.* (1960) and Ebihara *et al.* (2017), to keep consistency with the previous estimates for the historical storms (Miyake *et al.*, 2019; Usoskin *et al.*, 2023). We quantified the equatorward boundaries of the auroral emission regions by measuring their invariant latitude (ILAT) and tracing the footprints of the magnetic field lines where the auroral electrons precipitated, following McIlwain *et al.* (1961).

In the Southern Hemisphere, we have reconstructed the equatorward boundaries of the auroral oval down to 29.8° ILAT based on a Namibian report from Tivoli Astrofarm (S23°28′, E018°01′, −23.2° MLAT) with a naked eye auroral visibility up to 45°. This is somewhat followed by the Australian report from Bromfield Swamp (S17°22′, E145°33′, −24.5° MLAT) with a naked eye auroral visibility up to 20° and the New Caledonian report from Boulouparis (−26.5° MLAT) with a naked-eye auroral visibility of up to 20° that allow the reconstruction of the equatorward boundaries of the auroral oval down to 35.0° ILAT and 36.7° ILAT, respectively. In the Northern Hemisphere, a Mexican report from Chirimoyos (N23°26′, W105°48′, 31.0° MLAT) allowed us to reconstruct the equatorward boundaries of the auroral oval down to 35.5° ILAT based on the reported naked-eye auroral altitude of 60°. This is slightly more equatorward than what we can reconstruct from another Mexican report from El Peyote (N22°41′, W100°36′, 30.7° ILAT) with a reported naked-eye auroral altitude of 25°. Overall, our reconstructions in both hemispheres were reasonably consistent: 29.8° ILAT in the Southern Hemisphere and 35.5° ILAT in the Northern Hemisphere.

Caveats must be noted on these estimates here. Firstly, their naked-eye visibility is considerably different from the camera visibility (Appendix 1). Secondly, SAR arcs often reach farther towards the magnetic equator from the auroral ovals. It is certainly tempting to associate this considerable visibility stretch to the SAR arcs (Spogli *et al.*, 2024), whereas this association is conflicting with the analysis of the GOLD imagers (Karan *et al.*, 2024).

The SAR arcs appear monochromatically in reddish colouration, in contrast to the actual reddish auroral emissions (Kozyra *et al.*, 1997; Shiokawa *et al.*, 1997). Mendillo *et al.* (2016) reported that patches of emission at the oxygen green line (557.7 nm) are superimposed to the SAR arcs. After considering the possibility that precipitating ions gave rise to the oxygen red line, Rees and Deehr (1961) suggested that precipitation of low energy electrons could result in the emission at the oxygen green line, together with head conduction from the plasmapause-ring current interaction region, which could result in the SAR arcs. For now, we cannot exclude the possibility that the monochromatic reddish aurora witnessed at low latitudes was associated with the SAR arcs, as was also the case with historical auroral reports.

Therefore, we need to choose records with non-reddish colours or explicit structures (Appendix 2) to rule out SAR arcs and conservatively reconstruct the equatorward boundaries of the auroral oval to be free from contamination of SAR arcs. The reports that we used to reconstruct the equatorward boundaries of the auroral oval in both hemispheres included non-reddish colourations (pinkish and purplish components in Bromfield in Australia, and pinkish components in Chirimoyos in Mexico; see Appendix 2) and explicit structures (in Chirimoyos, Mexico and Tivoli Astrofarm in Namibia), as partially seen in Figure 14. Our reconstructions of the equatorward boundary of the auroral oval support analyses of the GOLD imager that suggested auroral stretch down to the southern tip of South Africa and South America (Karan *et al*., 2024), rather than the SAR arc hypothesis based on a model calculation (Spogli *et al*., 2024). Therefore, our reconstructions enhance the reliability of an extension of the equatorward boundary of the auroral oval during the May 2024 storm in comparison to the modelling results of Spogli *et al*. (2024).

## 8. Temporal and spatial variation of electron density (total electron content) in the ionosphere

In the previous sections, we elucidated the characteristics of temporal and spatial variations of the geomagnetic field, ionospheric convection, and location of the auroral oval during a geomagnetic storm. Considering the magnitude of geomagnetic field perturbations and ionospheric convection, a substantial electromagnetic energy transfer from the magnetosphere to the ionosphere is expected to produce severe positive and negative storms that cause the electron density in the ionosphere to increase or decrease. To confirm these signatures in the ionosphere during a geomagnetic storm, we analysed the global total electron content (TEC) data at a high spatial resolution of $0.5° \times 0.5°$ (geographic latitude × longitude).

Figure 16 shows the polar map of the ratio of the TEC difference (rTEC) (Shinbori *et al*., 2019) in the Northern Hemisphere in altitude adjustment-corrected geomagnetic (AACGM) coordinates (Section 4.2 in Laundal and Richmond (2017)) at 17:00 and 22:00 UT on May 10 and 18:00 UT on May 11, corresponding to the pre-storm phase, main, and recovery phases of this geomagnetic storms, respectively. During the pre-storm phase, the rTEC value was nearly zero across the entire region, except for the west coast of North America, exhibiting an MLAT range of 50–70°. This means that the spatial distribution of rTEC is geomagnetically quiet. During the main phase, the characteristic structure of rTEC appeared in the high- to low-latitude regions. Furthermore, the rTEC considerably increased to > 1.0 at low latitudes (20–50° MLAT) in the dusk sector (13–20 h MLT). These characteristics correspond to those of storm-enhanced density (SED) (Foster, 1993). Furthermore, the SED is associated with a spatially narrow enhancement of the rTEC, extending from the dayside to the nightside in the polar region. This phenomenon is called the tongue of ionisation (TOI) and is caused by plasma transportation from the dayside mid-latitude ionosphere into the polar cap owing to the enhanced convection electric fields during the main phase of a geomagnetic storm (David *et al.*, 2011; Foster, 1989, 1993). Shinbori *et al.* (2022) demonstrated the seasonal dependence of SED and TOI amplitudes as a function of the geomagnetic storm magnitude and based on a statistical analysis, they found that the SED and TOI amplitudes were maximal in winter and minimal in summer. However, despite the summer season, when most of the northern polar region is sunlit, a large rTEC enhancement related to the SED and TOI occurrences during the geomagnetic storms of 10–11 May 2024 was observed, which occurrence differed from those reported in previous studies. This implies that an intense convective electric field was imposed from the high-to-low latitudes of the ionosphere during the geomagnetic storm events. Furthermore, a narrow-band rTEC enhancement, corresponding to the extension of the auroral oval, was observed in the MLAT range of 45–55° on the nightside (22–03 h MLT), implying that the auroral oval extended to lower latitudes, as described in the previous section. Similar to the rTEC enhancement, a narrow-band region with decreased rTEC appeared in the MLAT range of 25–45°. This rTEC depletion represents a mid-latitude trough corresponding to the plasmapause location in the inner magnetosphere (Shinbori *et al.*, 2019). During the recovery phase, the global spatial distribution of the rTEC considerably decreased in the entire region, from high-to-low latitudes, reaching −0.75 on the dayside. This rTEC depletion corresponded to a negative storm caused by a decreasing electron density in the ionosphere owing to the enhancement of the recombination process associated with an increasing $N_2$ density. Thus, the extensive depletion of electron density in the ionosphere severely influences the propagation of the electromagnetic waves employed in satellite navigation and positioning.

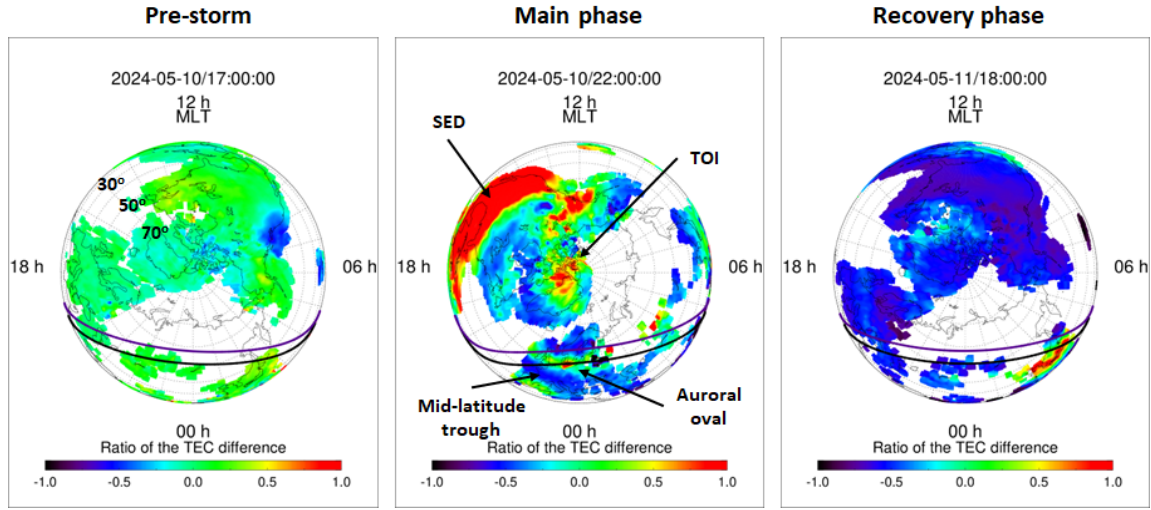

Figure 16: Polar maps of the TEC-difference ratio (rTEC) in the Northern Hemisphere represented in altitude adjustment-corrected geomagnetic (AACGM) coordinates during the pre-storm, main, and recovery phases. The colour bars indicate the rTEC from −1.0 to 1.0. The black and purple curves represent sunset terminators at altitudes of 105 and 300 km, respectively.

## 9. Summary and Discussion

This paper presented a flash report that reviewed the data from the May 2024 solar and geomagnetic storms. The area of the source AR 13664 rapidly increased from 113 to 2761 MSH from 1–9 May (Figures 1 and 2). This rapid growth led to an increase in the total energy of the nonlinear force-free field, potential magnetic field, and free magnetic energy, especially 7 May onwards, producing 12 X-class flares up to X8.7 between May 8 and 15 (Table 1 and Figure 2).

This AR was associated with at least ten halo CMEs. By evaluating their velocities, we could associate the CME launched at 22:24 UT on 8 May 2024 with the great geomagnetic storm that originated from the SSC at 17:05 UT on 10 May (Figure 4). Thus, the propagation time of the ICME was 42 h 41 min. This ICME collided with at least three preceding ICMEs and all of them stacked up before reaching Earth, leading to the detection of large IPS amplitudes. These signals are associated with the high-density region where fast-propagating ICMEs accumulate background solar wind and might support ICME–ICME interactions in interplanetary space.

One of these solar eruptions triggered a significant Forbush decrease in cosmic-ray intensity and a solar proton event. The ICMEs caused a relatively large Forbush decrease with the magnitude of

≥15% as seen in neutron monitors' count rate. Subsequently, both the polar neutron monitors and the GOES-16 satellite registered a notable cosmic-ray increase from ≈ 2:00–3:00 UT (Figure 6). This increase was confirmed to be a GLE and included as GLE #74 in the International GLE Database.

These ICME arrivals considerably compressed the magnetosphere. THEMIS measurements confirmed their crossings over the boundary of the magnetosphere at ≈ 8.24 $R_e$ at 17:05 UT (Figure 7), ≈ 8.51 $R_e$ at 17:05 UT (Figure 8), ≈ 5.04 $R_e$ at 19:12 UT (Figure 9), and ≈ 5.37 $R_e$ at 19:19 UT (Figure 10). Particularly, the observations of THEMIS-E and THEMIS-A during their crossings at 19:12 and 19:19 UT, respectively, indicated that the magnetopause underwent a significant inward motion beyond the geosynchronous orbit (6.6 $R_e$). Additionally, major loss processes for radiation belt particles occur during such great events (Turner *et al*., 2012). Moreover, these events have profound space-weather implications for geospace satellites because they can be exposed to energetic radiation belt particles that may partially or completely damage their electronic systems (Hands *et al.*, 2018). For example, if a satellite re-enters the magnetosphere and radiation belts when the magnetopause moves back out, the likelihood of damage is amplified if energetic-particle injections associated with rapid energetic-particle injections and wave-particle interactions increase (Baker *et al.*, 2017; Koons & Fennell, 2006). Therefore, satellites in geospace may have been exposed to harsh space-weather conditions during the geomagnetic storm of 10 May 2024.

The SSC at 17:05 UT upon shock arrival was observed at multiple observatories such as Kakioka (≈ 78 nT) and San Fernando (≈ 130 nT), indicating local enhancements in the duskside, especially in the summer hemisphere. This LT effect probably enhanced SSCs in the European sector.

The resultant geomagnetic storm developed to a minimum Dst and Dxt index values of −412 and −421.8 nT, respectively, at 2:00 UT on 11 May. In the provisional AE and AL indices, the auroral electrojet was intensified up to 4098 and −3797 nT during the storm main phase, respectively, and 3982 and −3797 nT during the storm recovery phase, respectively. However, the intensity in both indices reduced between the phases, probably because the auroral oval moved equatorward and remained away from the AE/AL observatories. However, their magnitudes should not be exaggerated as statistical studies have indicated that the return period of the Dst index is 5.7 years (Tsubouchi and Omura, 2007) and that of the AE index 1.49 years (Nakamura *et al*., 2015). It is possible that geomagnetic activity during Solar Cycle 24 was considerably lower than that during normal solar cycles.

Furthermore, the DP2 ionospheric current was significantly enhanced (Figure 14). This enhancement manifested as an intensification of the magnetospheric convection, which is highly likely to cause earthward penetration of the hot plasma originating in the plasma sheet, thereby enhancing the ring current.

This storm also induced considerable disturbances in the ionosphere (Figure 16), leading to an SED in the ionosphere, large rTECs in the lower MLATs (20–50°) on the duskside (13–20 MLT), and tongues of the ionosphere extending from the dayside to the nightside in the northern polar region, despite unfavourable seasonality. These findings indicated the occurrence of an intense convective electric field in the high-to-low MLATs in the ionosphere. These TEC records also confirmed rTEC enhancements in a narrow MLAT band of 45–55° on the nightside, confirming the equatorward extension of the auroral oval in accordance with the visual records. The rTEC measurements also revealed a plasmapause in the inner magnetosphere, characterised by a depletion in the 25–45° MLAT region and a considerably negative storm globally during the recovery phase.

Among the records collected since the beginning of the space age (1957 onwards), the min Dst of −412 nT is comparable to those of four of the five greatest geomagnetic storms, specifically those that occurred in September 1957 (min Dst = −427 nT), February 1958 (min Dst = −426 nT), July 1959 (min Dst = −429 nT), and November 2003 (min Dst = −422 nT), and significantly lower only than that of the extreme storm recorded in March 1989 (WDC for Geomagnetism at Kyoto *et al*., 2015; Riley *et al*., 2018). This value should be revised in the future because this real-time Dst index was developed using a method different from that for the final Dst index and extended only up to 2016 at the time of writing this manuscript. Additionally, this storm recorded the highest value on the Kp index (Kp = 9) (Matzka *et al*., 2021) since October 2003 (Yamazaki *et al*., 2024). Moreover, it is the second to the fifth greatest on the Hp30 index since 1985 (Yamazaki *et al*., 2024). In this regard, this geomagnetic storm was certainly large but not unique.

500 Years of Space Weather Storms with Aurora Visible at or Equatorward of 30° Magnetic Latitude

Figure 17: Spatial extents of naked-eye auroral visibility of great historical geomagnetic storms whose auroral visibilities have been reported as <|30|° MLAT. This figure is a modified version of those presented in Knipp *et al*. (2021) and Hayakawa *et al*. (2024).

During this storm, the auroral display was visible down to |23.2|° MLAT (Tivoli Astrofarm in Namibia) with the naked eye (Figure 14). This allowed us to reconstruct the equatorward boundaries of the auroral oval at an ILAT of 29.8°. In Figure 17, this storm is contextualised with auroral visibilities of historical storms, as modified from the previous presentations in Knipp *et al*. (2021) and Hayakawa *et al*. (2024). Based on naked-eye observations, this storm was comparable to the great storms of February 1958 and July 1959, for which the equatorward boundary of the auroral oval was reconstructed down to |23.3|° and |27.4|° MLAT, respectively (Hayakawa *et al*., 2023a, 2024).

Historical records indicate that the equatorward boundaries of auroral ovals have extended down to 33–40° ILAT during the greatest geomagnetic storms of the space age (Boteler, 2019; Hayakawa *et al*., 2023a, 2024) and 24–25° during extreme historical storms (Hayakawa *et al*., 2020, 2023b). Analysing modern naked-eye observations of auroral ovals during geomagnetic storms and comparing them with those of historical ones is crucial for better contextualising great storms.

However, our reconstruction was most likely an ultra-conservative estimate. The scientific community hosts large-scale citizen science portals (Macdonald *et al*., 2015; Kosar *et al*., 2018a; Wicklund, 2024), which can significantly increase the number of auroral records. The utility of these citizen science aurora datasets such as Aurorasaurus has already been confirmed through ground-truth validation of satellite observations and boundary estimations in a case study of the intense geomagnetic storm of March 2015 (Kosar *et al*., 2018b). In the near future, we will request detailed reports from individual contributors to document the temporal evolution (Figure 12), spatial evolution (Figure 15), and colouration of the reported auroral displays (Figure A2). These details will allow us to enhance the reconstruction of auroral emissions, their temporal/spatial evolution, and their equatorward auroral oval boundaries. Moreover, as this storm occurred in mid-May, the European and North American sectors had the disadvantages of longer daytime durations and better weather conditions. The timing of the storm peak suggests that the auroral visibility conditions were favourable in the Southern Hemisphere, particularly in South America. Therefore, we need to request for the recollection of this visual aurora by citizens, especially individual astronomers, in multiple languages. Additionally, outreach efforts to explain scientific phenomena in plain language are imperative for including and empowering the public, as exemplified by the Aurorasaurus blog.

Immediately after this storm, auroral photographs were uploaded from various regions worldwide. This significant geographical extent can be partially explained by the differences in the sensitivities of naked-eye observers and cameras. Today, we have access to significantly better optical instruments, such as cameras, than during past great space-weather events. Moreover, space-weather forecasts have allowed and motivated people to observe aurorae at the right time and direct cameras to the sky to figure out whether they can capture aurorae, as confirmed by several studies. Furthermore, Internet/SNS accessibility allows observers to swiftly exchange reports/photos, which motivates more people to search for visible aurorae. However, light pollution in urban areas hinders aurora visibility in the night sky (Smith, 2008; Falchi *et al.*, 2016). Therefore, witnesses of this auroral storm have three advantages and one disadvantage when documenting auroral extensions during great geomagnetic storms than those witnessed historical auroral storms.

This study is no more than a flash report conducted using provisional datasets, and more precise datasets must be accumulated and compared with the modelling results. Through further analyses, these solar and geomagnetic storm features may allow us to bridge our modern scientific knowledge with that of historical great solar and geomagnetic storms and better assess space-weather extremes.

**Data Availability**

We have acquired sunspot number and sunspot drawings from the SILSO and USET of the Royal Observatory of Belgium, the solar wind parameters and proton and X-ray fluence from the NOAA, the IPS data used from the ISEE, Nagoya University website https://stsw1.isee.nagoya-u.ac.jp/vlist/, neutron monitor data from the International GLE database, and geomagnetic Dst, AE, and AL indices from WDC and the location of north geomagnetic pole for Geomagnetism at Kyoto (https://wdc.kugi.kyoto-u.ac.jp/). We also acknowledge the Oulu Cosmic-Ray Station, the Neutron Monitor Database and individual neutron monitor station managers for maintaining the cosmic-ray measurement datasets over a long time period. The TEC data are available on the ISEE, Nagoya University website https://stdb2.isee.nagoya-u.ac.jp/GPS/GPS-TEC/. We thank Samuel Freeland for offering Event Archives of Solar Soft in the Lockheed Martin Solar & Astrophysics Laboratory.

**Acknowledgments**

This research was conducted under the financial support of JSPS Grant-in-Aids JP20H05643, JP21K13957, JP21H04492, and JP24H00022, the ISEE director's leadership fund for FYs 2021 -- 2024, the Young Leader Cultivation (YLC) programme of Nagoya University, Tokai Pathways to Global Excellence (Nagoya University) of the Strategic Professional Development Program for Young Researchers (MEXT), the young researcher units for the advancement of new and undeveloped fields in Nagoya University Program for Research Enhancement, and the NIHU Multidisciplinary Collaborative Research Projects NINJAL unit "Rediscovery of Citizen Science Culture in the Regions and Today". This study was supported by the Research Council of Finland (projects 330063 QUASARE and 354280 GERACLIS). This work was partially funded by the Horizon Europe program projects ALBATROS and SPEARHEAD. We acknowledge the support of the International Space Science Institute (Bern, Switzerland) and International Team No. 585 (REASSESS). IPS observations were made under the solar wind program of the Institute for Space-Earth Environmental Research, Nagoya University.

We thank Yuichi Otsuka, Michi Nishioka, and Septi Perwitasari for developing and sharing the TEC codes, Ilya Usoskin and Keith Ryden for their helpful discussions for the GLE #74, Takuya Tsugawa and Mamoru Ishii for their valuable advice on the ionospheric disturbances at that time., Elizabeth Macdonald for her helpful discussions on the citizen-science approach on the auroral records, Ravindra Desai for interpretations of the datasets of the CMEs and the solar wind, and Mirko Piersanti for his helpful discussions on the Italian geomagnetic measurements. We thank Tony Philipps, Martin Snow, Cristina Mandrini, Andrew Lewis, Hong-Jin Yang, Jacques van Delft, Kazue Nakagawa, and Ichiro Ota for helping us distribute the said survey for the local auroral visibility.

We thank Frédéric Desmoulins, Carlos Matos, Angelica D. Vazquez Sepulveda, Rene Saade, Shiori Yamada, Ryu Tamura, Trent Davis, Joel Weatherly, Tina Booth, Laura-May Abron, Landon Oxford, Leyton Riley, David Batchelor, Víctor R. Ruiz, Yuto Hoshino, Kate Green, Karl Krammes, Huili Chai, Horace A Smith, Akihiro Tamura, Edwin Rivera, Alan Viles, Jeff Vollin, Stella Stritch, Sam Deutsch, Robert Stuart, John Bradshaw, Shigeki Tomita, M. L. Couprie, Ida Kraševec, Jesse Wall, Rosenberg Róbert, Tim Martin, Jimmie Strouhal, Tyler McLain, Greg Redfern, Michael Borman, Evan Saltman, Miriah Shadara, Helen Spillane, Jamie McBean, Graham Whittington, Clark Austin, Drew Medlin, Laura Lockhart, Larkyn Timmerman, Ricardo Muracciole, Laura Lockhart, Joël Van Quathem, Francisco Martinez Nieto, Gentrit Zenuni, Genna Chiaro, Joseph Jiacinto, Paul D. Maley, Julia Sumerling, Denis Martínez, Brandon Flores, Yutaka Kagaya, Gerard Kelly Edgar Castro, Edy

## Appendix 1: Spatial Extents of the Auroral Visibility and the Auroral Oval: Instrumental Observations

Upon this storm, considerable observers managed to photograph the auroral displays by their instruments such as digital cameras and smartphones. Such camera observations managed to capture the auroral visibility and auroral oval further equatorward. Good examples can be easily obtained from Japanese cases. As shown in Figure A1, the aurora was photographed at least down to Koumi (N36°04, E138°24, 28.1° MLAT). More photographs on social media show auroral visibility at lower latitudes, but they require further careful validation (*e.g.*, Meel and Vishwakarma, 2020). In contrast, the naked eye visibility is confirmed only down to Hiranai (N41°01′, E140°53′, 33.2° MLAT). The Hiranai photographs captured aurora, on the basis of their evident ray structure, while the Koumi photograph might have captured SAR arcs, based on the appearance of monochromatic reddish colour (Figure A1).

The Hiranai report showed striking differences in the spatial auroral extents: by the naked eye, up to 20° in elevation, and by optical instruments, up to 70° in elevation. Following the procedure described in Section 7, this study allowed us to reconstruct the equatorward boundary of the auroral oval at Hiranai, down to 43.0° ILAT and 36.8° ILAT, for the naked-eye-based and camera-based estimates, respectively. The equatorward boundary looks significantly decent in contrast to our reconstructions in Section 6, as Japanese observers were under broad daylight around the storm peak (≈ 2 UT), only managed to see the storm recovery phase, and probably missed the most dramatic part of the auroral display.

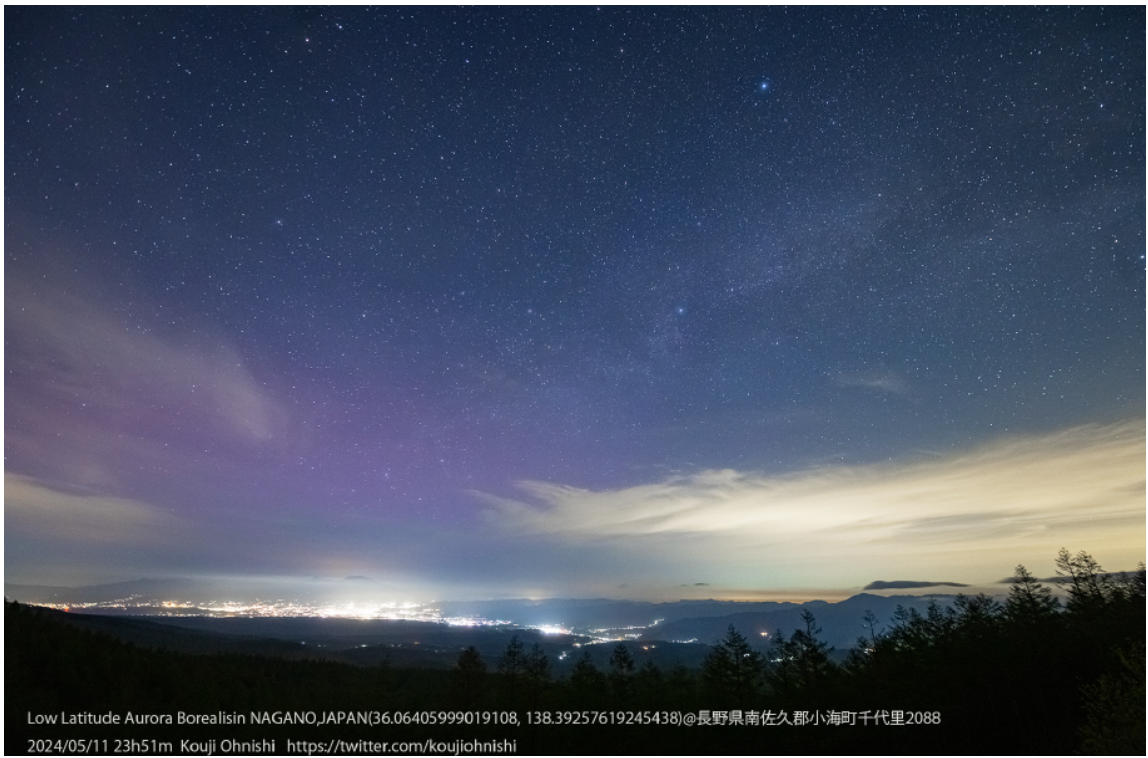

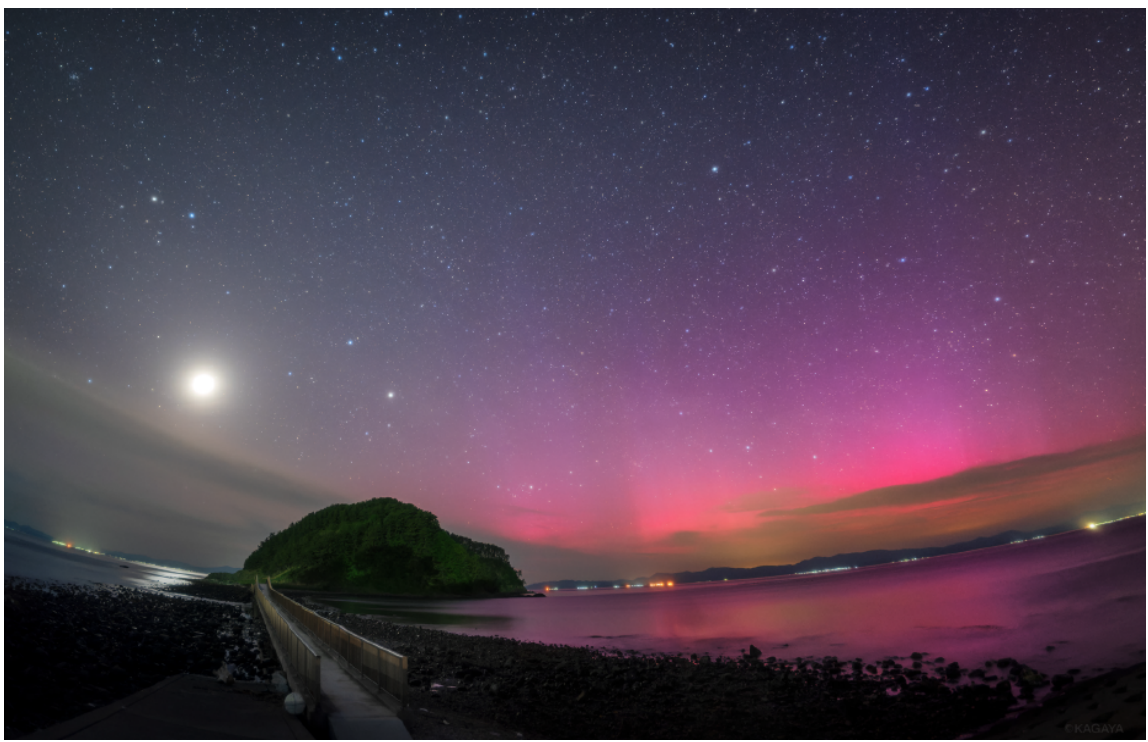

Figure A1: Examples of auroral photographs from Japan, from the top to the bottom: (a) Koumi in Japan with Canon EOS R5 (© Kouji Ohnishi) and (b) Hiranai in Japan with SONY α7R IV (© Yutaka Kagaya). Kouji Ohnishi confirmed that the aurora was not visible to the naked eye. Yutaka

Kagaya confirmed the spatial extent of the auroral display up to 20° in elevation by the naked eye, and up to 70° in elevation by optical instruments.

Similar cases have been reported throughout the world. An Indian auroral report from Hanle in India (N32°28′, E078°34′, 24.3° MLAT) described the auroral altitude therein as 5° by the naked-eye and 20° by camera. Similarly, the New Caledonian report from Boulouparis described auroral altitude as 20° by the naked eye and 30° by a camera. At Tivoli Astrofarm in Namibia (S23°28′, E018°01′, −23.2° MLAT), aurora was confirmed up to 45° by the naked eye and up to 55° by the camera. Subsequently, its internal structure was confirmed to be up to 20° by photographic analysis (Figure 14). Notably, these examples occurred at different times during the storm, as shown in Figure 12.

Overall, in the Northern Hemisphere, the aurora was photographed as far equatorward as San Juan Sacatepéquez in Guatemala (N14°43′, W090°39′, 23.4° MLAT) subvisually without naked eye visibility. In the Southern Hemisphere, the aurora was photographed as far equatorward as Big Mitchell Reserve in Australia (S16°48′, E145°21′, −23.0° MLAT). This estimate is probably ultra conservative. Independent citizen-science datasets accommodate potential auroral photographs from further equatorward, down to São Paulo in Brazil (S23°48′, W046°38′, −15.2° MLAT) and Lima (S12°05′, W076°52′, −2.9° MLAT) in the Southern Hemisphere and Ad-Dāḫilīyah in Oman (N22°52′, E057°32′, 16.8° MLAT) and Calabarzon in the Philippines (N14°06′, E121°05′, 5.2° MLAT) in the Northern Hemisphere, according to Wicklund (2024). However, as Wicklund (2024) does not provide further details, we need to carefully assess their reliability on the basis of their source reports.

These geographical extensions are significantly different from those in the naked eye visibility. The instrumental detectability should be of course somewhat variable up to the instrumental set ups such as exposure time. This variation is particularly tricky, as each observer had their own instrument such as digital cameras and smartphones without standardisations unlike well-prepared campaign observations. Such record characteristics have made calibration and homogenisation of their photographs extremely difficult.

The equatorward extent of the visibility with camera is different from that with naked eyes. This discrepancy is primarily attributed to the different sensitivity. The sensitivity of the camera is much higher than that of naked eyes. We could suggest three possible scenarios for its provenance as detectability difference in (1) vertical extents, (2) latitudinal extents, or (3) their combinations. For Scenario (1), the aurora at high altitudes could have been detected only by the cameras, so that the aurora was visible at lower latitudes. For this scenario, the equatorward boundary is assumed to be the same. For Scenario (2), the equatorward boundary of the auroral oval was fairly broad in latitude, and the aurora extended further equatorward diffusively. According to the DMSP satellite observations, the equatorward boundary of the electron precipitation is broad in latitude during large magnetic storms (Shiokawa *et al.*, 1997; Ebihara *et al.*, 2017). Precipitations of electrons with low energy and low flux could cause auroral emission further equatorward, but the auroral emission could be too dim to be visible to the naked eye observations. For this scenario, the altitude distribution of the aurora is assumed to be the same. It is of course possible that both were the case, as expected in Scenario (3). Further quantitative assessments are needed here.

For the moment, we accommodate as wide error margins as possible. It is empirically known that the maximum brightness of the reddish component of the aurora (at 630.0 nm) is located around 220 km in altitude, and it can range from 200 to 300 km (Solomon *et al.*, 1988). On the basis of spectra of precipitating electrons observed at mid latitudes near the storm maximum of the large magnetic storm of 13-14 March 1989, the volume emission rate of 630.0 nm peaks at about 270 km altitudes (Ebihara *et al.*, 2017). At 400 km altitude, the volume emission rate of 630.0 nm is an order of magnitude lower than that at the peak altitude. At 500 km altitude, the volume emission rate is much lower than that at the peak altitude. From this consideration, we set the altitudes of the upper limit of the aurora identified by camera to be 300, 400, 500, and 800 km. In this case, the Namibian report from Tivoli Astrofarm ($-23.2°$ MLAT), which has a local altitude with instrumental detectability of 55° above the local horizon, allows us to derive the equatorward boundaries of the auroral oval as, at least, 27.6°, 28.9°, 30.1°, and 33.3° ILAT, respectively. Alternatively, the Australian report from Big Mitchell Reserve ($-16.8°$ MLAT), which has a local altitude with instrumental detectability of 50° above the local horizon, allows us to derive the equatorward boundaries of the auroral oval as 22.4°, 24.0°, 25.4°, and 29.1° ILAT, respectively. Further detailed assessments are obviously needed with parallel case studies.

## Appendix 2: Spatial distribution of the Auroral Activity

Figure 15 shows a graphical summary of the temporal and spatial evolution of auroral visibility during the May 2024 storm, showing variations in the maximum auroral brightness. Based on the real-time Dst index, the auroral visibility below |40|° MLAT was concentrated in the late main phase and early recovery phase of this storm. One interesting account is a subvisual auroral observation from a ship in the South Atlantic Ocean (S32°00′, E17°00′, 31.5° MLAT) that lasted from 18:15 to 18:45 UT on 10 May. This aurora was photographed shortly after a SSC, despite its occurrence at lower latitudes. A similar case was observed during the March 1989 storm (Boteler, 2019).

This figure also shows the temporal and spatial evolution of the maximum auroral brightness at each site. We gathered and compared citizen contributors' self classification and descriptions on the maximum brightness in total illumination with an IBC reference (Chamberlain, 1961, p. 124) with an extension: (0) not visible by the naked eye; (1) Comparable to Milky Way; (2) comparable to thin moonlit cirrus clouds; (3) comparable to moonlit cumulus clouds; comparable to the full moon; and (4) brighter than any such references. This figure shows that the maximum brightness was higher in higher latitudes and lower at lower latitudes. This, in turn, shows gradual decreases in auroral brightness in lower MLATs and partially explains the smaller geographical extent of the naked-eye auroral visibility than that by instrumental observations.

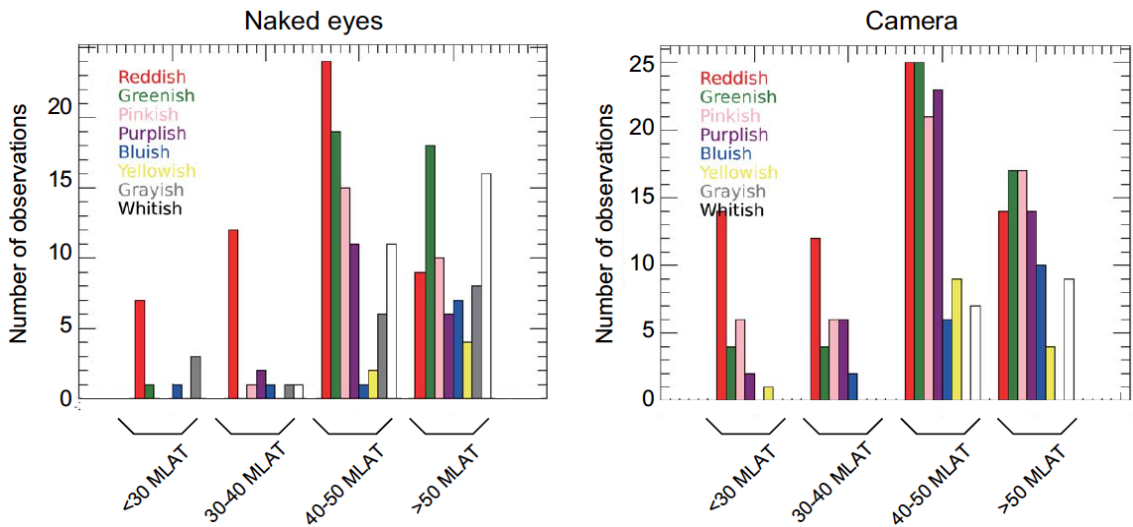

Figure A2: Amounts of reported auroral colourations in each MLAT ranges by naked eye (left) and by camera (right), showing (1) < 30° MLAT, (2) 30° − 40° MLAT, (3) 40° − 50° MLAT, and (4) > 50° MLAT, from left to right.

Figure A2 shows the number of auroral colourations reported in each MLAT range by the naked eye and the camera. A considerable number of reports show reddish and greenish colouration. These colours are typically associated with oxygen emissions at 630.0 nm (reddish) and 557.7 nm (greenish), owing to the precipitation of electrons. Reddish aurorae can more easily be seen than green aurorae by camera. Greenish aurorae cannot be spotted by eye as far south, generally because they occur at lower altitude. The yellowish colour is most likely caused by a mixture of reddish and greenish colours. The number of reports with greenish colouration was comparable to that with reddish colouration at > 40° MLATs, whereas it significantly decreased below 40° MLAT by naked eye observation. For camera observations, a greenish colouration was reported below 40° MLAT. Purplish and pinkish colours have also been reported below 40° MLAT. They are typically associated with nitrogen emissions, including 427.8 nm for $N_2^+$. Although bands of $N_2^+$ are known to coexist in the reddish auroral arc (Rees and Deehr, 1961), this may challenge the conventional wisdom that a low-latitude aurora is dominated by a reddish colour. For the purplish auroras that appear at high altitudes, the energy of the precipitating electrons is not necessarily high. The colouration tempts us to associate it with sunlit aurorae (Hunten, 2003), based on the similarity with the past auroral displays with bluish/purplish colouration during the great geomagnetic storms in 1909 and 1958 (Hayakawa *et al*., 2019, 2022a). The purplish aurora appeared to be dominant over the reddish one in some of the pictures, as exemplified with Víctor R. Ruiz's photograph in Mogán of Gran Canaria (Figure 14). The dominance of the purplish aurorae should be investigated in future studies, including different response bands of cameras and cell phones. It should be noted that the greenish colouration should have been seen in different colours, such as greyish, by naked eye observation.